\documentclass[12pt,tightenlines,eqsecnum,floats,showpacs,nofootinbib,amsmath,amssymb,aps,prd]{revtex4}

\begin{document}

\title{General Relativity and Gravitation: A Centennial Perspective}
\author{Abhay Ashtekar${}^1$, Beverly K. Berger${}^2$, James Isenberg${}^3$,
and Malcolm A. H. MacCallum${}^4$} 
\affiliation{1. Institute for Gravitation \& the Cosmos, and Physics
Department, Penn State, University Park, PA 16802, U.S.A.\\
2. 2131 Chateau PL, Livermore, CA 94550, USA\\
3. Department of Mathematics, University of Oregon, Eugene, OR
97403-1222, USA\\
4. School of Mathematical Sciences, Queen Mary University of London,
Mile End Road, London E1 4NS, U.K. }

\begin{abstract}

To commemorate the 100th anniversary of general relativity, the
International Society on General Relativity and Gravitation (ISGRG)
commissioned a Centennial Volume, edited by the authors of this
article. We jointly wrote introductions to the four Parts of the
Volume which are collected here. Our goal is to provide a bird's eye
view of the advances that have been made especially during the last 35
years, i.e., since the publication of volumes commemorating
Einstein's 100th birthday. The article also serves as a brief
preview of the 12 invited chapters that contain in-depth reviews of
these advances. The volume will be published by Cambridge University
Press and released in June 2015 at a Centennial conference sponsored
by ISGRG and the Topical Group of Gravitation of the American
Physical Society.

\end{abstract}


 \maketitle



\section{Introduction}
\label{s1}

The discovery of general relativity by Albert Einstein 100 years ago
was quickly recognized as a supreme triumph of the human intellect.
To paraphrase Hermann Weyl, \emph{wider expanses and greater depths
were suddenly exposed to the searching eye of knowledge, regions of
which there was not even an inkling.} For 8 years, Einstein had been
consumed by the tension between Newtonian gravity and the spacetime
structure of special relativity. At first no one had an appreciation
for his passion. Indeed, ``as an older friend,'' Max Planck advised
him against this pursuit, ``for, in the first place you will not
succeed, and even if you succeed, no one will believe you.''
Fortunately Einstein persisted and discovered a theory that
represents an unprecedented combination of mathematical elegance,
conceptual depth and observational success. For over 25 centuries before this discovery, spacetime had been a stage on which the dynamics of matter unfolded. Suddenly the stage joined the troupe of actors. In subsequent decades
new aspects of this revolutionary paradigm continued to emerge. It
was found that the entire universe is undergoing an expansion.
Spacetime regions can get so warped that even light can be trapped
in them. Ripples of spacetime curvature can carry detailed imprints
of cosmic explosions in the distant reaches of the universe. A
century has now passed since Einstein's discovery and yet every
researcher who studies general relativity in a serious manner
continues to be enchanted by its magic.

The International Society on General Relativity and Gravitation
commissioned a volume to celebrate a century of successive triumphs
of general relativity as it expanded its scientific reach. It
contains 12 Chapters, divided into four Parts, highlighting the
advances that have occurred during the last 3-4 decades, roughly
since the publication of the 1979 volumes celebrating the centennial
of Einstein's birth. During this period, general relativity and
gravitational science have moved steadily toward the center of
physics, astrophysics and cosmology, and have also contributed to
major advances in geometrical analysis, computational science,
quantum physics and several areas of technology. The next two
decades should be even more exciting as new observations from
gravitational wave detectors and astronomical missions open
unforeseen vistas in our understanding of the cosmos. The volume
provides a vivid record of this voyage.

The organization of the volume is as follows:\\

\noindent PART I: EINSTEIN'S TRIUMPH

\noindent Chapter 1:\\
   	100 Years of General Relativity\\
\emph{George F. R. Ellis} \medskip

\noindent Chapter 2:\\
   	Was Einstein Right?\\
\emph{Clifford Will} \medskip

\noindent Chapter 3:\\
Relativistic Astrophysics\\
\emph{John Friedman, Peter Schneider, Ramesh Narayan, Jeffrey E. McClintock, Peter M\'{e}sz\'{a}ros, and Martin Rees}\medskip

\noindent Chapter 4:\\
Cosmology\\
\emph{ David Wands, Misao Sasaki, Eiichiro Kamatsu, Roy Maartens and
Malcolm A. H. MacCallum}\\

\noindent{PART II: NEW WINDOW ON THE UNIVERSE}

\noindent Chapter 5:\\
Receiving Gravitational Waves\\
\emph{ Beverly K. Berger, Karsten Danzmann, Gabriela Gonzalez,
Andrea Lommen, Guido Mueller, Albrecht Ruediger, and William Joseph
Weber}
\medskip

\noindent Chapter 6:\\
Sources of Gravitational Waves: Theory and Observations\\
\emph{Alessandra Buonanno and B. Sathyaprakash}\\

\noindent PART III: GRAVITY IS GEOMETRY AFTER ALL

\noindent Chapter 7:\\
Probing Strong Field Gravity Through Numerical Simulations\\
\emph{Frans Pretorius, Mattew Choptuik and Luis Lehner}
\medskip

\noindent Chapter 8:\\
Initial Data and the Einstein Constraint Equations\\
\emph{ Gregory Galloway, Pengzi Miao and Richard Schoen}
\medskip

\noindent Chapter 9:\\
Global Behavior of Solutions to Einstein's Equations\\
\emph{ Stefanos Aretakis, James Isenberg, Vincent Moncrief and Igor
Rodnianski}\\
\vfill\goodbreak

\noindent PART IV: BEYOND EINSTEIN

\noindent Chapter 10:\\
Quantum Fields in Curved Space-times\\
\emph{Stefan Hollands and Robert Wald}
\medskip

\noindent Chapter 11:\\
From General Relativity to Quantum Gravity\\
\emph{Abhay Ashtekar, Martin Reuter and Carlo Rovelli}
\medskip

\noindent Chapter 12:\\
Quantum Gravity via Supersymmetry and Holography\\
\emph{Henriette Elvang and Gary Horowitz}\\

The goal of these Chapters is two-fold. First, beginning researchers
should be able to use them as introductions to various areas of
gravitational science. Second, more advanced researchers should be
able to use the Chapters that are outside the area of their
immediate expertise as overviews of the current status of those
subjects. Since the scope of gravitational science has widened
considerably in recent years, the volume should be useful not only
to specialists in general relativity but also to researchers in
related areas.

In this article we have collected the detailed introductions to the
four Parts. They summarize the main advances and offer a short tour
of the more detailed material that can be found in the Chapters that
follow. The introductions also contain illustrative examples of
outstanding important open issues, and outline important
developments that could \emph{not} be included in individual
Chapters where the authors had the difficult task of covering many
developments in a limited space. We hope that this material will
provide a global perspective on developments in the four main areas
of gravitational science.

\section{Part I: Einstein's Triumph}
\label{s2}

Recent media attention to the centenary of the outbreak of the First
World War (WWI) reminds us that it was against this backdrop that
Einstein, a Swiss citizen, announced the revolutionary theory of
general relativity (GR).  The war affected the theory's
dissemination. Eddington's report introducing GR to the
English-speaking world \cite{Edd18} relied on information from de
Sitter in neutral Holland. Inevitably, the theory's adherents were
caught up in the conflict, most notably Karl Schwarzschild, who died
in 1916 while serving on the Russian front.

In 1915 Einstein was already a decade on from his {\em annus
mirabilis} of 1905, in which he had announced the theory of special
relativity, explained the already well-observed photoelectric effect
as due to quantization of light (a vital step towards quantum
theory), and explained Brownian motion assuming the reality of
atoms, an explanation experimentally confirmed by Perrin in 1908.
The second of these three great papers won him the 1921 Nobel prize
-- and they were not all he published that year! For example, he
gave the famous $E=mc^2$ equation, which later gave the basis of
nuclear fusion and fission (whence Einstein's intervention in the
development of atom bombs). Fusion in particular explained how stars
could hold themselves up against gravity as long as they do. So
Einstein had already triumphed well before 1915.

However, he was aware that his work left an awkwardly unresolved
question -- the need for a theory of gravity compatible with special
relativity that agreed with Newton's theory in an appropriate limit.
Here we will not recount Einstein's intellectual development of
general relativity, which resolved that problem, nor describe the
interactions with friends and colleagues which helped him find the
right formulation. Those are covered by some good histories of
science, and biographies of Einstein, as well as his own writings.

The theory's prediction of light-bending confirmed to good accuracy
\cite{Har79} by the UK's 1919 eclipse expedition led by
Eddington%
\footnote{How Eddington, a Quaker, while preparing for this
expedition, avoided being sent to work on the land as a
conscientious objector, is itself an interesting WWI story.}
and Crommelin, brought Einstein to the attention of the general
public, in particular through the famous headline in the New York
Times of November 9th. From then on, he increasingly came to be seen
as the personification of scientific genius.

Why then are we calling this first Part of our centennial book
``Einstein's triumph''? GR had already triumphed by 1919.

The triumph since 1919 lies in GR's ever increasing relevance and
importance, shown in particular by the number and range of
applications to real world observations and applications, from
terrestrial use in satellite navigation systems to considerations of
cosmology on the largest scales. Moreover the different applications
are now interwoven, for example in the relevance of black holes in
cosmology and the use of pulsars and compact relativistic stars in
strong field tests of the theory. This Part of the book outlines
that progress.

As Ellis describes in Chapter 1, the starting points for many later
confirmations were laid in the early years of the theory: the
Schwarzschild solution, leading to solar system tests and black hole
theory; light-bending, which grew into gravitational lensing; and
the Friedmann(-Lema\^{i}tre-Robertson-Walker: FLRW) solutions, basic
in cosmology. Moreover, several confirmations relate to the three
``classical tests'': gravitational redshift, the anomaly in the
perihelion advance of Mercury as computed from Newtonian theory%
\footnote{One may note that the anomalous part is
43${}^{\prime\prime}$ per century in a total of around
5000${}^{\prime\prime}$ per century.},
and light-bending: for example, the analysis of GPS (the Global
Positioning System), the study of the binary and double pulsars, and
the use of microlensing to detect exoplanets. The theory remains the
most nonlinear of the theories of physics, prompting development in
analytic and numerical technique.

Classical differential geometry as studied in introductory courses
(and as briefly outlined by Ellis) is adequate to discuss the
starting points of those developments. But they soon require also
the proper understanding of global structure and thus of
singularities and asymptotics, for example in understanding the
Schwarzschild solution, black holes and the energy carried away by
gravitational radiation. This increasing sophistication was
reflected in the best-selling text of Hawking and Ellis
\cite{HawEll73}, and further developments are described in Part
Three of this book.

Much of the development of GR has come in the last half century. For
its first 50 years, a time when quantum theory was making big
advances, one could argue that GR remained an intellectual ornament
with only some limited applications in astronomy. Even its relevance
to cosmology was debatable, because Hubble's erroneous distance
scale led to a conflict between the geologically known age of the
Earth and the age of the universe in a FLRW model, prompting the
range of alternative explanations for this discrepancy described in
Bondi's book \cite{Bon60}. While the notion of a stagnant phase is
rather belied by the many significant papers from this time which
have deservedly been included in the ``Golden Oldies'' series of the
\emph{General Relativity and Gravitation} journal, some of them
cited by Ellis, it was certainly a less dynamic period than the
following 50 years of GR.

The changes have been partly due to the already mentioned increasing
mathematical sophistication among theoretical physicists. Taub's use
of symmetry groups \cite{Tau51}, and Petrov's algebraic
classification of the Weyl tensor \cite{Pet54}  were crucial steps
forward made in the 1950s. The geometric concepts of connection and
curvature have become fundamental in modern gauge theories. Progress
in the theory of differential equations has given a firm basis to
the idea that GR is like other physical theories in that initial
configuration and motion determine the future evolution. The
generating techniques for stationary axisymmetric systems used to
obtain exact solutions%
\footnote{The construction and interpretation of exact solutions are
topics not covered by this book, as they are well covered by
\cite{SteKraMac03} and \cite{GriPod09} and references therein. In
particular we do not consider some important techniques used in
those areas, such as computer algebra and the application of local
spacetime invariants.}
relate to modern work on integrable systems. Further developments in
such areas are reflected in Chapter 1 and Part Three of this book.

Another important step was introducing the theory of the matter
content within FLRW models. This enabled the understanding of the
formation of the chemical elements, by combining the Big Bang and
stellar nucleosyntheses, the provision of evidence that there were
only three types of neutrino, and the prediction of the Cosmic
Microwave Background (CMB).

Progress has  depended even more on advances in technology and
measurement technique. The first example was the revision of
Hubble's distance scale in 1952 by Baade, using the 200 inch Palomar
telescope commissioned in 1950. This led to increasing belief in the
FLRW models, a belief eventually cemented by the 1965 observations
of the CMB, which themselves arose from developments in microwave
communications technology.

The 1957 launch of the first artificial satellite, Sputnik,
intensified the need for detailed calculation of orbital effects in
satellite motion, in order to very accurately plan satellite
projects. Such work \cite{Bru91} was undertaken for both the US and
USSR programs and was the first practical use of GR.

Radio astronomy, by showing source counts inconsistent with the
alternative Steady State theory, had provided important evidence for
FLRW models. It also, combined with optical observations, led to
the discovery of quasars%
\footnote{3C48 was identified in 1960 and 3C273's redshift was found
in 1963.}
which prompted Lynden-Bell to propose that they were powered by
black holes \cite{Lyn69}: the importance black holes have
subsequently assumed in our understanding of astronomy and cosmology
is described by Narayan and McClintock in Chapter 3. Radio astronomy
also discovered the pulsars, announced in 1968, which gave extra
impetus to the already developing study of relativistic stars,
discussed by Friedman in Chapter 3.

The reality of gravitational waves in the theory, which had been
debated earlier, was finally clarified in the work of Bondi et al in
1959 \cite{BonPirRob59}. The binary and double pulsar observations,
described in Chapter 2, united the understanding of compact objects
and gravitational waves to provide the first strong field tests of
GR.

The exquisite precision now achieved in practical and observational
areas of GR has made use of the development of very high precision
atomic clocks and of the burgeoning of electronics since the
invention of the transistor in 1947. Satellite-borne telescopes in
several wavebands, computers of all scales from the largest (used in
numerical relativity) to mobile devices (e.g.\ in GPS receivers),
CCD devices (based of course on the photoelectric effect), and
lasers (in terrestrial gravitational wave detectors -- also used,
for example, in determining the exact position of the moon) have all
played major roles in the observations and experiments described in
the following four Chapters (and in the later parts of the book).

There were fundamental aspects of gravity (e.g.\ the E\"otv\"os
effect) which could be and were tested on Earth, but until the 1970s
the focus was on the ``classical tests'', complemented by the time
delay measurements for satellites. Dicke initiated a more systematic
analysis of the equivalence principle and its tests, as described in
Chapter 2. Thorne, Will and others then developed other frameworks,
notably the PPN framework, which could encompass other types of
tests. While the application of these ideas still relied on solar
system and terrestrial tests, these became much more precise and
involved much new technology (e.g.\ laser ranging to the moon,
superconducting gravimeters on the ground, use of atomic traps and
atomic clocks in terrestrial and satellite experiments), and pinned
the parameters of the PPN framework down with high precision.

Tests outside the solar system consisted of the understanding of
compact stars such as white dwarfs, and supernova remnants, and of
cosmology (for which there was only an incomplete understanding, for
reasons described below), but did not lead to new precise
constraints on the theory. That changed with the discovery and
observations of the (first) binary pulsar, and still further with
the several now known, including the double pulsar. These give some
of the most precise measurements in physics (although, perhaps
surprisingly, the Newtonian constant of gravitation, $G$, remains
the least accurately known of the fundamental constants of nature).

It is notable that the understanding of pulsars not only required GR
(because of the strong fields) but also entailed the simultaneous
use of quantum theory and GR (because only by taking into account
quantum theory could one have adequate equations of state to model
white dwarfs and neutron stars). These types of compact objects, and
black holes, are now the starting points for the calculation of
gravitational wave sources described in Part Two.

Relativistic astrophysics then developed in a number of directions
(see Chapter 3). Numerical simulations gave much more detail on
relativistic stars, their properties, stability and evolution. A
whole new sub-discipline of black hole astrophysics came into being,
concerned with the environments of black holes, especially (for
stellar size black holes) accretion from neighboring stars, and (for
supermassive black holes) accretion, nearby orbits and tidal capture
of stars. The improved understanding enabled us to be rather certain
not only that there really are black holes in the Universe, but that
they are very common.

A further direction described in Chapter 3 came about with the
discovery and increasingly detailed observations of gamma ray
bursts. Both their long and short varieties turned out to require
models of relativistic sources, as described by M\'esz\'aros and
Rees. It is interesting that there is a link with the gravitational
wave detectors described in Part Two, in that the absence of
gravitational waves from GRB 070201 showed that, if it had a compact
binary progenitor, then that progenitor had to be behind rather than
in M31 \cite{Hur08}.

While the standard FLRW models used up to 1980 or so did very well
in describing the observed isotropy and homogeneity of the universe,
and explaining the evolution of the matter content which led to
formation of the chemical elements and the prediction of the CMB,
they failed to explain the single most obvious fact about the
Universe, namely that it has a highly non-uniform density. Naturally
occurring thermal fluctuations and their evolution could not give
large enough variations. The inflationary paradigm, introduced by
Guth in 1981 \cite{Gut81}, altered that radically by providing
reasons for a nearly flat spectrum of density fluctuations at a time
sufficiently early in the universe for the subsequent linear and
nonlinear phases of evolution to produce the observed structures we
see. The theory is described in detail by Sasaki in Chapter 4.

The resulting standard model has been compared with a range of very
high precision observations, notably those of the CMB, the baryon
acoustic oscillations (BAO) and the magnitude-redshift relation for
supernovae (relating distances and expansion velocities in the
Universe). These, especially the CMB observations, have generated
the title ``precision cosmology'', which, as Komatsu emphasizes in
Chapter 4, requires precision theory as well as precision
observation. That precision in theory consists of very detailed
consideration of perturbations of the FLRW models and of light
propagation in perturbed models, enabling the link between the
conditions produced by inflation (or some alternative to inflation
providing suitable initial conditions) and the present-day
observations. What is remarkable is the fine detail of those initial
conditions that one can infer from observation.

To some degree, the role of GR has disappeared in the large volume
of literature related to CMB, BAO and supernova, and other
observations, as almost all of it uses the FLRW models and their
linearized perturbations, and may even use Newtonian analyses in key steps. Wands and Maartens remind us, in their introduction to Chapter 4, that GR in fact still has a crucial role
to play, even in precision cosmology where its effects may be
considerably larger than the very small error bars in the
observations, and the correlations described in Chapter 3 imply it
also has a role to play in structure formation below the scales
tested by the CMB. Moreover it is essential in testing the
robustness of the assumptions of the concordance model, a further
topic discussed in Chapter 4.

What can we expect in the future? The results of Planck's B-mode
polarization%
\footnote{There are two characteristic patterns of polarization
alignments expected in the CMB. The E-mode is like that of the
electric field round a charge and the B-mode like that of magnetic
field round a current. Instances of these modes, with varying
amplitudes and centered at random locations, will be superposed in
the actual observations. For more details see Chapter 4.}
measurements, confirming or contradicting those of BICEP2 and
POLARBEAR (both of these being outlined in Chapter 4), may be
published before this book. In 2015 Advanced LIGO will begin taking
data (see Part Two). If such advanced gravitational wave detectors
see the expected gravitational wave sources, we will have a new
window for testing GR (but if no such sources are seen, that may be
due only to poor astrophysical predictions). In the past, when new
windows on the universe have opened, new and unforeseen phenomena
have been found \cite{Har84}; it would not be surprising if this
happens again. Beyond that there are a plethora of new instruments
being built or planned to study the sky in electromagnetic wavebands
from low frequency radio to $\gamma$-rays: the chances of convincing
funding agencies to support such work have probably been
substantially enhanced by the spectacular results of recent past
projects.

Gravitational lensing by galaxies seemed to surprise many when first
found in 1979, even if it should not have. Now such lensing, along
with its stellar size counterpart, have become tools for astronomy,
used for example to infer the distribution of mass within galaxies,
the distribution of dark matter, the properties of distant galaxies,
and the presence of new exoplanets. Recently, magnification due to
microlensing was used to determine properties of a binary system
containing a white dwarf and a Sun-like star \cite{KruAgo14}.

We stress again that the galactic and intergalactic application is
just one of the instances where different aspects of GR come
together -- here lensing and cosmological models.

Although the greatest challenge for GR may lie in finding and
testing a good enough theory of quantum gravity, as discussed in
Part Four, there are still challenges at the classical level.
Cosmology provides the greatest of these, since its standard model
requires three forms of matter -- the inflaton, dark matter and dark
energy -- which have not been, and perhaps cannot be, observed in
terrestrial laboratories, and whose properties are modeled only in
simple and incomplete ways. It would of course be ironic if the
triumph of GR in cosmology were to turn to disaster because the only
way to deal with those apparently-required three forms of matter
were to adopt a modified theory of gravity, but other explanations
seem much more likely.

The inflaton is postulated as a way to produce the nearly flat
spectrum of fluctuations required as initial data from which
acoustic oscillations produce the observed CMB power spectrum. While
the assumptions of inflation may be questionable, it is, as already
mentioned, remarkably successful in producing the right distribution
of fluctuations on present day scales above 150 Mpc or so (a scale
much larger than that of individual galaxies). Moreover the recent
detection of B-modes claimed by BICEP2 is consistent with inflation
theory's prediction of polarization due to gravitational waves. A
definitive detection of such polarization would provide indirect
evidence on quantum gravity and the quantum/classical
correspondence, in that the theory assumes the quantum fluctuations
of the inflationary era become classical.

The evidence for dark matter is rather securely based on
observations at scales where a Newtonian approximation is good
enough to show that not all the mass is visible, such as
observations of galactic rotation curves and the distribution of
X-ray emitting gas in clusters. It provides 25-30\% of the critical
energy density of the Universe, itself now known to be very close to
the actual energy density (see Chapter 4). This was known before the
more precise CMB and baryon acoustic oscillation (BAO) measurements
\cite{ColEll97}. Additional evidence for dark matter  has been provided by comparing
the distribution of mass in colliding galaxies, as shown by its
lensing effects, with the mass distribution of the visible gas.
However, a change in the gravity theory might provide an explanation
for these observations not requiring dark matter, though
as yet no satisfactory such theory has been proposed.%

The inference of the existence of dark energy is even more dependent
on GR, in particular on the theory of perturbed FLRW models (see
Chapter 4): it comes from the magnitude-redshift relation for
supernovae (relating distance and expansion velocity of the
Universe), CMB and BAO data. Attempted explanations within GR not
requiring a new form of matter (in which we include the cosmological
constant) have used both large and small scale inhomogeneities (see
Chapter 4), or may arise from the astrophysics of supernovae and
their environments. Or we may be able to pin down the properties of
dark energy in some way independent of FLRW models, and thereby
provide a further triumph for the predictions of GR.

Obtaining information about the three so far unobserved constituents
of the standard model may not come from GR itself. But we would
certainly like a better understanding of inhomogeneities and their
back reaction and impact on light propagation. The evidence of
correlations of galactic properties with central black hole masses
suggests we also need to know much more about the messy non-linear
processes of galaxy and star formation and their interaction with
the nonlinearities of GR.

Despite these lacunae, which may offer opportunities for future
breakthroughs, when taken together the following four Chapters
illustrate very well the staggering extent of the triumph of
Einstein's 1915 proposal of the theory of General Relativity.

\section{Part II: New window on the universe}
\label{s3}

Gravitational waves provide an opportunity to observe the universe
in a completely new way but also give rise to an enormous challenge
to take advantage of this opportunity. When Einstein first found
wave solutions in linearized general relativity and derived the
quadrupole formula, it became clear that a laboratory experiment to
produce and detect gravitational waves was impossible while it was
also clear that any gravitational wave signals produced
astronomically were too weak to be detected on earth with the
instruments available or thought possible at that time. Nearly 100
years later, we are at the confluence of fundamental science and
technology that will soon open this new window.

Several lines of development were required to make the search for
gravitational waves realistic.  Despite the early recognition by
Einstein that linearized gravity had wave solutions, the physical
reality of gravitational waves remained in dispute for many decades.
The reason for this was the absence of formalisms able to separate
physical degrees of freedom in the field equations from coordinate
(gauge) effects. A well known, striking example was Einstein's
conviction that the Einstein-Rosen cylindrical waves
\cite{Einstein:1937} were not physical and furthermore that the
character of this exact solution proved that there were no physical
gravitational waves in the full theory. While Einstein retrieved the
correct interpretation in the nick of time \cite{Kennefick05}, the
question remained unsettled until correct, gauge invariant
formulations of the problem were developed. The first of these, from
Bondi's group \cite{Bondi:1957dt,Pirani:1956wr,Bondi:1959}, used the
``news function'' to demonstrate that, far from the source, one
could quantify the energy carried away by gravitational waves.
Further developments in understanding equations of motion, gauge
freedom, and other methods to identify gravitational waves in the
background spacetime led to approximation methods with greater
precision and broader application than the original linear waves
\cite{isaacson68a,Isaacson68b}. In addition, the first half-century
of general relativity saw the physically relevant exact solutions of
Schwarzschild \cite{Schwarzschild16} and, much later, Kerr
\cite{Kerr63}. In the late 1930s, Oppenheimer and Volkoff
\cite{Oppenheimer39a} and Oppenheimer and Snyder
\cite{Oppenheimer39b} studied spherically symmetric gravitational
collapse that predicted the formation of neutron stars and black
holes respectively even though both categories were considered
highly unlikely as real objects for several decades thereafter.

Meanwhile, astronomical technology developed beyond the optical
frequency band. Radio astronomy coupled to optical followup led to
the discoveries of quasars in 1963 and pulsars in 1967 to usher in
the age of relativistic astrophysics---not to mention the discovery
of the cosmic microwave background (CMB) in 1965. Suddenly, the
universe became much more interesting for general relativity. X-ray
detectors aboard rockets found bright point sources outside the
solar system, possibly indicating a more violent universe than
previously suspected. Supernovae could now be better studied and
served as the driver for early general relativistic computer
simulations \cite{May66}. At approximately the same time as physical
systems in the strong-gravity regime became plausible, theoretical
developments made it clear that gravitational waves could carry
energy and thus could be physically relevant. At some point,
improved treatment of the two-body problem in general relativity led
to a number of recalculations of Einstein's quadrupole formula,
engendering a rather vitriolic and long reigning dispute over the
coefficient. It became clear later (in the 1980s) that consistent
approximations yielded Einstein's results \cite{Kennefick07}.

The possibility that detectable gravitational wave sources might
exist and the firm theoretical foundation that such waves were a
prediction of general relativity inspired Joseph Weber to build a
detector to search for them \cite{Weber57}. His first detector, an
aluminum cylinder (bar) with piezoelectric readout is described in
\cite{Weber67,Weber69}. He quickly realized that coincident
detection---i.e., two or more bars---was an essential ingredient.
This first generation of bar detectors reached a sensitivity to
gravitational wave strain amplitudes of $10^{-16}$ by the end of the
1960s \cite{saulson05}. Weber's electrifying claim of the detection
of signals associated with the center of our galaxy \cite{Weber69}
inspired a worldwide effort leading to one or more bar detectors
operating in China, France, Germany, Italy, Japan, UK, USA, and USSR
\cite{Aguiar11,saulson05}.  Unfortunately, even though the
subsequent experiments of this first generation of bars were more
sensitive than Weber's, they failed to confirm his detection. In
addition to bar development, plans were made to use the earth
\cite{Levine72}, the moon \cite{Giganti77}, and a number of
spacecraft \cite{Armstrong06} as gravitational wave detectors.  By
the first decade of this century, the largest bars had reached the
size of 3 meters and the mass of more than 2 tons, were operated at
cryogenic temperatures as low as 0.1 K to minimize thermal noise,
and achieved a strain sensitivity of $\approx$ $10^{-21} /
\sqrt{Hz}$ at 900 Hz. Prototypes of novel designs for resonant mass
detectors were developed in this period with at least one prototype
currently under construction \cite{Aguiar11}.

In the mid 1970s, interferometric detectors were first considered
both for the ground and for space. Chapter 5 focuses on highlights
of the current status of such instruments. The first prototypes were
built by Forward (at Hughes) and later by Weiss (at MIT) and Drever
(at Glasgow and Caltech) as well as elsewhere in the world
\cite{saulson05}. The laboratory-scale interferometers had
sensitivities far below those of the contemporaneous bars. However,
the interferometers could be scaled up in contrast to bars (see
\cite{saulsonbook}) but would require facility-class dimensions and
funding to achieve sensitivities that were plausible for detection.

We note here that gravitational wave detection pushes precision
technology and related theory and thus can yield important
scientific spinoffs. An early example is quantum non-demolition. The
broad idea is to overcome the standard quantum limit by arranging to
decrease the uncertainty in one variable by increasing it in the
complementary variable. This was first proposed for bar detectors by
Braginsky et al \cite{Braginsky77} and refined for interferometers
by Caves \cite{Caves81}. The concept has been studied extensively in
atomic, molecular, and optical physics since then but is most
spectacularly implemented on the kilometer-scale interferometer
GEO\,600 \cite{Schnabel11} and was tested successfully on one of the
LIGO instruments \cite{Aasi13b}. It is likely to become a near-term
upgrade for the second generation interferometers Advanced LIGO and
Advanced Virgo.

Meanwhile, in the early 1970s, the Uhuru X-ray satellite revealed
the X-ray sky for the first time (where brief rocket flights had
only given hints). Stellar mass binaries, including Cyg X-1, a
strong black hole candidate, as well as a number of active galactic
nuclei, were revealed and could be correlated with radio and optical
sources. It became possible to argue that potential gravitational
wave sources were abundant.

This time period also saw the first steps toward binary black hole
simulations by Smarr and collaborators \cite{Smarr79}. However, the
landmark discovery of this decade was the binary pulsar PSR1913+16
by Hulse and Taylor \cite{Hulse75}. Not only was this the first
known system containing two neutron stars, but the orbital
parameters made it sufficiently general-relativistic to strongly
constrain theories of gravity (see the discussion in Chapter 2).
Even more exciting, the period decay due to energy loss by emission
of gravitational waves was measurable. Because this system is clean
(no extraneous effects from tides or matter accretion), it provided
incontrovertible experimental evidence of the existence of
gravitational waves and also of the correctness of the quadrupole
formula (see the discussion in Chapter 6). As a side note, this
validation of the quadrupole formula killed off the last vestiges of
debate about the coefficient. In addition, the known binary neutron
star systems are precursors of gravitational wave sources for
ground-based detectors---and, in fact, are the only ones that
demonstrably exist. Although supernovae in our galaxy may be
detectable sources, the details of nonaxisymmetric collapse are not
yet known sufficiently well to predict their strength. What this
means is that the orbit decay of binary neutron stars will
eventually lead to the final inspiral and merger phase that produces
strong gravitational waves at frequencies of up to approximately one
kHz (see Chapter 6). Because the time to merger for the known
systems is significantly less than the age of the universe, one has
reason to hope that such systems in the merger phase occur with
sufficient frequency to be a target for human detection if
sufficiently sensitive instruments could be built. As more (but
still a small number) such systems were discovered and stellar
evolution became better understood, event rate estimates became
possible \cite{Abadie:2010cfa}. Chapter 6 focuses on the
gravitational-wave signatures of these and other proposed sources.
To obtain the correct waveform requires either careful (or even
rigorous) approximations or robust numerical simulations.

While the best understood source for ground-based interferometric
detectors has been observed electromagnetically in its precursor
phase, actual known, named objects, namely white dwarf binaries, are
accessible to space-based detectors which when constructed and
launched will operate at the much lower frequencies characterizing
such systems. Examples are given in Chapter 6.

A byproduct of the expansion of astronomy into non-optical detection
was the identification of systems of stellar mass and up to $10^9$
M$_{sun}$ which could be black holes. Over the years, the black hole
hypothesis has become the favored explanation for these systems.
Black hole binary systems, if they exist, would be an obvious target
for gravitational wave detection since in a clean environment of no
accretion discs they would emit no electromagnetic signals. Hawking
had proved that a merging binary black hole could emit up to $29\%$
of its mass-energy in gravitational waves \cite{Hawking71}. But what
was the actual number? To answer this question required numerical
simulations in three spatial dimensions (3D). While the 2D (head on
collision) problem was solved reasonably well by Smarr and
collaborators \cite{Smarr79} in the 1970s, the 3D problem proved
astonishingly difficult until it was finally solved in 2005
\cite{Pretorius2005a,Baker2006a,Campanelli2006a} (see Chapter 7).

The growth of relativistic astrophysics made the scientific case for
development of gravitational wave detection programs compelling.
Although many bar groups left the field after their failure to
confirm Weber's claims, several remained and continued to improve
their instruments \cite{saulson05}. Nonetheless, though the
interferometric prototypes built in the 1970s and 1980s were less
sensitive than the bars of their era, it became clear that
interferometers were the only way to achieve the needed sensitivity
in the frequency range accessible from the ground. Efforts were
begun in the US, Germany, UK, France, Italy, and Japan to seek
funding to build the kilometer-scale instruments that would be
necessary to detect a strong gravitational wave event from the Virgo
cluster (see Chapter 5). This distance was chosen as a criterion
because it is necessary to probe at least that far to have sufficient
likelihood for a detectable event (see Chapter 6). Remarkably, given
the novelty and scale (both physically and financially) of the
proposed technology, two kilometer-scale projects, LIGO in the US
and Virgo, a French/Italian collaboration in Italy, with 4 km and 3
km arm-lengths respectively, were funded and built. Construction
took place in the 1990s and data taking by some of the instruments,
as well as by the shorter but nearly kilometer-scale GEO\,600 and
TAMA, began in about 2000 (see Chapter 5).

Meanwhile, building on earlier efforts with data from prototype
detectors \cite{Allen99}, programs to analyze the data coming from
LIGO, Virgo, GEO\,600, and TAMA were initiated. The time series
$h(t)$ for the gravitational wave strain was analyzed either by
matched filtering (comparing to templates) for binary neutron stars
or as unmodeled bursts (see Chapters 5 and 6). Note that the former
required the substantial computational power provided by advancing
technology to become feasible. Toward the end of the 2000s,
astrophysically interesting upper limits were obtained, the
technological tour-de-force of design sensitivity was achieved, and
bar sensitivity was exceeded \cite{Abadie12}.

In 2010, an upper limit for burst events was published, based on
observations by a world-wide network of bars \cite{Astone10}. While
the limit itself was not interesting astrophysically, the effort
demonstrated the advantages of international collaboration in this
field. In addition, the technology to rotate the LSU bar was
developed and implemented. This is interesting because a
cross-correlation, stochastic measurement with the LIGO Livingston
instrument allowed regular reorientation of the bar to impose a
periodicity on any signal present since the allowed orientations
included a direction where there should be no correlation
\cite{Abbott07}.

Starting in the 1980s, space-based concepts were proposed and
evaluated leading to adoption of LISA (first envisioned by Bender
and Faller) as a joint R\&D project of NASA and ESA in 1998. The
astronomical community became engaged, possibly due to the existence
of the white-dwarf-binary calibration sources. Regrettably, in
contrast to the ground-based detectors supported through physics
programs, the astronomy-centered space programs in Europe and the US
never moved LISA into a launchable status. One issue was the alleged
riskiness of the technology. While the level of precision needed for
LISA is six orders of magnitude less than for, e.g., LIGO due to
LISA's proposed arm-lengths of millions of kilometers (see Chapter
5), the picometer level of precision required is still daunting for
space-qualified missions and several necessary technologies are new.
It is hoped that the often postponed but now firmly scheduled LISA
Pathfinder will demonstrate the feasibility of LISA and LISA-like
missions (see Chapter 5).

In the past decade or so, increasing interest has arisen in taking
advantage of the remarkable timing precision and stability
associated with millisecond pulsars to use the monitoring of pulse
arrival times to tease out signals from passing gravitational waves
\cite{Hobbs10}. These methods are effective at nanohertz frequencies
and are sensitive to signals from supermassive binary black holes
and from a stochastic background. In principle, the ability to
detect a gravitational wave signal increases with increasing
monitoring time (with a power that depends on the type of source).
At this time, it is not clear that the systematic errors are
completely understood. Nevertheless, there is a significant chance
that pulsar timing arrays will detect gravitational waves within
this decade (see Chapter 5).

Of course, we should not leave out the imprint of gravitational
waves on the cosmic microwave background (see Chapter 3). In
particular, at certain angular scales, relic gravitational waves
produced in the early universe leave a curl-like tensor imprint, the
so-called B-modes, on the cosmic microwave background. If the signal
is sufficiently strong, with the ratio of tensor-to-scalar
polarization amplitude $r$ large enough to be currently observable,
the gravitational waves are likely to have been produced by
cosmological graviton production and amplified by inflation. A
recent report of such a detection with  $r \approx .2$ by BICEP2 may
have been premature since the measured effect could be caused by
foreground dust or have a significant dust component \cite{BICEP14}.
Whether or not this possible detection of gravitational waves is
confirmed, the discussion centered on the observation emphasizes
that direct or indirect detections of gravitational waves reveal
information about the universe and its contents that are
inaccessible with electromagnetic radiation, cosmic rays, or even
neutrinos.

A number of different technologies have been applied to the
different programs for gravitational wave detection. These are not
in competition but rather are complementary in that they search for
gravitational waves in different frequency bands. As with
electromagnetic emitters, gravitational waves at different
frequencies arise from different sources. In simple terms, the
frequency depends inversely on a power of the mass (see Chapter 6).
At the highest frequencies of 10 Hz - 10$^4$ Hz accessible from the
ground, the predominant sources are expected to be ``chirp signals''
from binary neutron stars or binary black holes with masses of order
1-300 M$_{sun}$, ``monochromatic'' signals from non-axisymmetric
rotating neutron stars, correlations between the noise from two
detectors that characterize a cosmological or astrophysical
stochastic background, or unmodeled bursts from supernovae in our
galaxy or other sources. The ground-based detectors certainly have
the potential to identify the engines of gamma ray bursts
\cite{Sathya09}.  Future generations may be able to probe the
equation of state of nuclear matter, discover intermediate mass
black hole binaries, and reveal as yet unsuspected new phenomena.
Space-based observations are sensitive to binary white dwarfs and
binary neutron stars in the early inspiral phase and binary black
holes in the mass range of $10^5$ to $10^7$ M$_{sun}$. Perhaps more
interestingly for general relativity, they can observe the inspiral
of a stellar mass black hole around a supermassive black hole. If
the gravitational waves from such sources can be observed with
sufficient precision \cite{Sathya09}, detailed tests of general
relativity can be made. One is the definitive ``smoking gun''
detection showing that the central object is a black hole (see
Chapter 6). Another allows one to obtain an interesting bound on the
graviton mass through a distortion of the phasing of gravitational
waves in a binary black hole inspiral \cite{Will14}. Finally, pulsar
timing arrays can search for gravitational wave signals from
supermassive binary black holes that are suspected to exist from
their electromagnetic properties, as well as ones that are so far
unsuspected. Once gravitational waves are detected, it would be
natural to expect that the experience of the early radio, X-ray, and
$\gamma$-ray telescopes will be repeated with the discovery of
completely new and unexpected phenomena to, once again, widen our
horizons in unforeseen directions.

The next two chapters explore the sources and technologies that
comprise gravitational wave science. They emphasize the growth in
our knowledge about the sources through observations, theory, and
simulations, including the state of our knowledge on their numbers
(or number density) in the universe, and their gravitational wave
signatures. These are then the targets for decades of technological
development for detectors on earth and in space with the former on
the verge of recording their first signals.

When looking back over a century of general relativity, we see that
gravitational waves not only represent a future transformative
discovery but have also, over the past 30 years or so, changed the
nature of the field of gravitational physics. General relativity has
gone from a discipline with few practitioners exploring apparently
arcane (but often seminal) mathematical properties of the theory to
big science with international collaborations yielding publications
with nearly 1000 authors whose instruments require an investment on
the order of \$1 billion (US). A further transformation will occur
with the first detections by the second generation Advanced LIGO and
Advanced Virgo instruments,  independent of the precise nature of
the sources. Unlike the cosmic microwave background measurements or
pulsar timing arrays that piggyback on instruments developed for
other purposes, the instrumentation developed for gravitational-wave
detection on the ground and in space represents a completely new way
to study the universe. It is likely that the discoveries soon to be
made by these advanced detectors will ``shake loose'' the future of
the field by accelerating the timeline for the first space-based
gravitational wave detectors and by encouraging investment in third
generation instruments such as the proposed Einstein Telescope
\cite{Punturo10}.

\section{Part III: Gravity is Geometry, after all}
\label{s4}

Einstein's general relativity is a mathematically beautiful
application of geometric ideas to gravitational physics. Motion is
determined by geodesics in spacetime, tidal effects between physical
bodies can be read directly from the curvature of that spacetime,
and the curvature is closely tied to matter and its motion in
spacetime. When proposed in 1915, general relativity was a
completely new way to think about physical phenomena, based on the
geometry of curved spacetimes that was largely unknown to
physicists.

While the geometric nature of Einstein's theory is beautiful and
conceptually simple, the fundamental working structure of the theory
as a system of partial differential equations (PDEs) is much more
complex. Einstein's equations are not easily categorized as
wave-like or potential-like or heat-like, and they are pervasively
nonlinear. Hence, despite the great interest in general relativity,
mathematical progress in studying Einstein's equations (beyond the
discovery of a small collection of explicit solutions with lots of
symmetry) was quite slow for a number of years.

This changed significantly in the 1950s with the appearance of
Choquet--Bruhat's proof that the Einstein equations can be treated
as a well-posed Cauchy problem \cite{CB52}. The long term effects of
this work have been profound: Mathematically, it has led to the
present status of Einstein's equations as one of the most
interesting and important systems in PDE  theory and in geometrical
analysis. Physically, the well-posedness of the Cauchy problem for
the Einstein equations has led directly to our present ability to
numerically simulate (with remarkable accuracy) solutions of these
equations which model a wide range of novel phenomena in the strong
field regime.

The Cauchy formulation of general relativity splits the problem of
solving Einstein's equations, and studying the behavior of these
solutions, into two equally important tasks: First, one finds an
initial data set---a ``snapshot" of the gravitational field and its
rate of change---which satisfies the \emph{Einstein constraint
equations}, which are essentially four of the ten Einstein field
equations. Then, using the rest of the equations, one evolves the
gravitational fields forward and backward in time, thereby obtaining
the spacetime and its geometry. The first of these tasks, working
with initial data and the constraint equations, is the focus of
Chapter 8. The second task, involving evolution is the subject of
Chapter 9. Chapter 7 discusses the ideas and methods which are used
to carry out both of these tasks numerically, as well as some of the
novel insights that have been obtained from this numerical work.

It is not surprising for the field equations of a physical theory to
include constraint equations. Maxwell's equations, for example,
include the constraints $\nabla \cdot B= 0$ and $\nabla \cdot E= 4
\pi \rho$. However, the Einstein constraints (see Chapter 8) are
much harder to handle than their Maxwellian counterparts, so
considerable mathematical research has gone into working with them
over the years.

The approach which has been most successfully used over the years
for constructing and analyzing solutions of the constraint equations
is the \emph{conformal method}. Based on the work of Lichnerowicz
\cite{L44} and York \cite{Y72}, the conformal method splits the
initial data into ``seed data", which can be freely chosen, and
``determined data", which one  obtains if possible by solving the
constraint equations with the seed data specified (see Chapter 8 for
details). The goal of the conformal method is two-fold: to obtain an
effective parametrization  of the ``degrees of freedom" of the
gravitational field, and to construct initial data sets which
incorporate the physics of interest.

For initial data with constant mean curvature (CMC), the conformal
method works very well. This is true for data on compact manifolds,
for asymptotically Euclidean (AE) data, and for asymptotically
hyperbolic (AH) data. It also works very well for data sets with
sufficiently small mean curvature (near-CMC), so long as there are
no conformal Killing fields present. However, more generally, much
less is known. There are special classes of non-CMC seed data for
which the conformal method has been shown to work \cite{HNT}, but
there are others for which it appears to behave quite badly
\cite{Max11}. While to date these latter classes are quite
restricted, the signs of trouble are clear: It may well be that the
conformal method is effective for CMC and near-CMC initial data, but
is ineffective more generally.

Are there modifications of the conformal method which might get
around these looming difficulties? We first note that while the
\emph{conformal thin sandwich method} \cite{Y99} is a popular
alternative to the conformal method among numerical relativists,
Maxwell has shown \cite{M14a} that these two methods are
mathematically equivalent---they succeed or fail together. Thus, one
expects that more substantial modifications may be necessary to
obtain a method which works for all solutions of the constraint
equations. Maxwell \cite{M14b} has proposed one possible direction
for such modifications, in terms of his ``drift" vector treatment of
the mean curvature in the seed data, but much more work is needed to
determine if this or any other modification is likely to be
successful.

While the conformal method has to date been the dominant tool used
for constructing and analyzing solutions of the constraint
equations, there are other approaches available. Most notable are
the \emph{gluing techniques}, which are designed to combine two or
more known solutions of the constraints into a single one. Roughly
speaking, there are two types of gluing techniques which have proven
to be effective. \emph{Connected sum} gluing starts with a pair of
smooth solutions of the constraints, chooses a point in each, and
produces a new  solution of the constraints on the connected sum
manifold%
\footnote{Roughly speaking, the connected sum of a pair of manifolds
is obtained by removing a neighborhood of a point from each, and
then adding a tube which connects the manifolds where the neighborhoods have been removed.}
which is identical to the original solutions away from  the chosen
gluing points.  This gluing technique \cite{IMP02, CIP} works for
any pair of initial data sets---compact, AE, or AH---so long as a
certain non degeneracy condition (see \cite{CIP}) holds at the
gluing points. It also works for most coupled--in matter source
fields. It has been used to prove a number of interesting results,
including that there exist maximally--extended globally hyperbolic
solutions of the constraints  containing \emph{no} CMC Cauchy
surfaces.

The other type of gluing was introduced in 2000 in the landmark work
of Corvino \cite{C00}, in which it is shown that for any
time--symmetric initial data set which solves the constraints and is
asymptotically Euclidean in a suitable sense, and for any bounded
open interior region  $W$,  there is a new solution of the
constraints which is identical to the original in $W$, and is
identical to a subset of the Schwarzschild initial data outside of
some bounded set containing $W$. The surprise here is that quite
general interior gravitational configurations can be smoothly glued
to a Schwarzschild exterior. Equally surprisingly, Corvino and
Schoen show in \cite{CS06} (see Chapter 8) that this result extends
to non--time--symmetric initial data, with the exterior region being
Kerr rather than Schwarzschild.  Based on this work, one can show
that there is a large class of asymptotically flat solutions of the
Einstein equations which admit the conformal compactification (and
the corresponding $\mathcal{I}$-structure) of the sort proposed by 
Penrose for studying asymptotically flat solutions.

Very recently, a further surprising variant of Corvino--Schoen type
gluing has been developed by Carlotto and Schoen. They show
\cite{CS14} that, just as one can construct a solution of the
constraints by gluing an arbitrary interior region across an annular
transition region to an exterior Schwarzschild or Kerr region, one
can also do so by gluing an arbitrary solid conical region
(stretching out to spatial infinity) across a transition region to a
region of Minkowski initial data  with a conical region removed.
Moreover, one can do this without significantly changing the ADM
mass.  Both the Carlotto--Schoen conical  gluing and the
Corvino--Schoen annular gluing \cite{CCI11}, can be used to produce
N--body initial data sets of specific design which satisfy the
constraints. The evolutions of such N--body initial data sets have
not yet been studied.

In addition to the conformal method and the gluing techniques, there
is one other procedure which has been used to produce solutions of
the constraints. Designed to construct solutions with fixed interior
boundaries, Bartnik's ``quasispherical method" \cite{B93} chooses
coordinates in such a way that one of the constraints takes the form
of a parabolic (heat--type) equation with the radial coordinate as
``time". Specifying data on the interior boundary, one ``evolves"
outward towards $r \rightarrow \infty$. Originally proposed as a
tool for studying the Bartnik quasilocal mass, the quasispherical
method has been used as well for constructing dynamical horizon
initial data sets \cite{BI06}. It would be useful to explore
possible further applications.

The development of procedures for constructing solutions of the
constraint equations is only one part of the broad scale of research
focussed on initial data sets and the constraints. Another very
important area aims to understand the energy, momentum and angular
momentum of isolated gravitational systems, both for the system as a
whole and for specified subregions. Such notions, well understood in
Newtonian physics, are not obvious in general relativity, especially
for non--isolated subregions. Globally, for a data set which is
asymptotically Euclidean, the ADM formalism  \cite{ADM}  provides
definitions which are correct from the  Hamiltonian perspective, and
are theoretically useful. In particular, the global ADM energy
$E_{ADM}$ and the  global ADM momentum $P_{ADM}$ are both crucial
for the statement and proof of two of the major results of
mathematical relativity: the \emph{Positive Mass Theorem} and the
\emph{Penrose Inequality Theorem}. As discussed in Chapter 8, the
first of these states that for every AE initial data set satisfying
the constraints, the corresponding $E_{ADM}$ and $P_{ADM}$ satisfy
the inequality $E_{ADM} \ge |P_{ADM}|$, with equality holding if and
only if the initial data set generates Minkowski spacetime. The
Penrose Inequality Theorem, which thus far has been proven only for
time--symmetric  initial data sets, states that for time-symmetric
AE solutions of the constraints  the ADM mass
$\mu_{ADM}:=\sqrt{E^2_{ADM}-|P_{ADM}|^2}$ satisfies the inequality
$\mu_{ADM}\ge \sqrt{\frac{A}{16\pi}}$, where $A$ is the surface area
of the outermost horizon in the data set.

The proofs of both of these theorems---the first by Schoen and
Yau \cite{SY79} and then independently by Witten \cite{W81} during
the late 1970s, the second by Huisken and Ilmanen \cite{HI01} and
then independently by Bray \cite{B01} at the turn of the
millennium---were major triumphs of geometric analysis. To a large
extent, they showed  mathematicians working in geometric analysis
that general relativity could be a  fertile source of mathematically
deep---and solvable---problems. Indeed, formulating and proving a
version of the Penrose Inequality Conjecture for
non--time--symmetric initial data is an outstanding open problem in
mathematical relativity and geometric analysis.

While the definitions, major results, and major open questions
regarding global quantities like $E_{ADM}$ are agreed upon by most
relativists, much less is settled for the localized quantities.
There is agreement regarding the properties that a ``quasi-local
mass" should have. Yet, as discussed in Chapter 8, there is a wide
variety of possible definitions, including the Brown--York mass, the
Liu--Yau mass, the Wang--Yau mass, the Hawking mass, and the Bartnik
mass. In addition, there are definitions of quasi--local mass
relying on isolated horizons and on dynamical horizons. All have
their virtues and their difficulties, and their studies motivate a
variety of challenging problems in geometric analysis.

One of the geometric structures which arose as a useful tool for the
Schoen--Yau proof of the Positive Mass Theorem is the marginally
outer trapped surface or MOTS. In a spacetime, an embedded
two--dimensional spacelike surface $\Xi$ is a MOTS if one of the two
families of future--directed null geodesics orthogonal to $\Xi$ has
vanishing expansion everywhere along $\Xi$. Such behavior is closely
tied to that of an apparent horizon, and consequently (through
Hawking--Penrose--type singularity theorems) to the development of a
black hole. One also notes that if $\Xi$ is contained in a
time--symmetric initial data set, then it must be a minimal surface
in that Riemannian manifold.

Given the strong tie between MOTS and black holes, and between MOTS
and minimal surfaces, it is not surprising that the analysis of
MOTS---conditions for their existence in a given spacetime,
conditions for their stability, restrictions on their allowed
geometry---has proven to be very rich. Chapter 8 covers these topics
extensively, including an outline of a very nice MOTS--based
simplification of the proof of the Positive Mass Theorem
\cite{EHLS}.

From a physical point of view, the analysis of initial data sets and
solutions of the Einstein constraint equations is all about
specifying physically interesting initial configurations of
gravitational systems, understanding the properties of these
systems, and identifying the degrees of freedom of the gravitational
field if possible. Correspondingly, from a physical point of view
the analysis of the evolution equations is all about determining how
(according to general relativity) such initial configurations evolve
in time: Do they form black holes? Do they form singularities? If
they produce gravitational radiation, what forms might the radiation
take for different initial configurations?

Choquet--Bruhat's well-posedness theorem shows that (in suitable
coordinates) Einstein's field equations constitute a (nonlinear)
hyperbolic PDE system. However, there are two very important ways in
which the Einstein system differs from other geometrically-based
hyperbolic PDE systems such as the Yang--Mills equations or  wave
maps on Minkowski spacetime. First, while the Yang--Mills and wave
map examples have a fixed background spacetime which can be used to
define long--time existence, singularities, and energy functionals
unambiguously, with Einstein's theory the spacetime evolves along
with the solution, so these concepts are much harder  to pin down.
Second, since general relativity uses solutions of Einstein's
equations to model (classical) physical phenomena, there is physical
interest in individual solutions and their evolution; for
Yang--Mills and wave maps this is not the case.

These two distinguishing characteristics of Einstein's equations
have had two important consequences. The lack of a fixed background
has made it difficult until recently to use some of the standard
ideas and tools of nonlinear hyperbolic PDE analysis. The physical
interest in individual solutions has made the pursuit of numerical
simulations of solutions a very important part of general relativity
since the 1970s.

While there was little progress in either  the numerical or the
hyperbolic PDE analysis of Einstein's equations in the two decades
following Choquet--Bruhat's theorem, important insights were
obtained regarding  singularities (see \cite{Ger}). It was known
from the study of explicit solutions that singularities (in the
sense of unbounded curvature) \emph{can} develop, but it was not
known if this was just an artifact of the symmetries of solutions
such as FLRW and Schwarzschild. The Hawking--Penrose ``singularity theorems", which
were proven using the properties of geodesic congruences in curved
spacetimes, did show that if one identifies geodesic incompleteness
with singularities, then singularities occur in large classes of
solutions. However, in making this identification, one obscures the
issue of the physical meaning of a singularity occurring in a given
spacetime. For example, in Taub--NUT spacetimes, geodesic
incompleteness marks the breakdown of causality (with closed
timelike paths appearing), while in FLRW spacetimes, geodesic
incompleteness occurs because of unbounded tidal curvature.

Restated as the \emph{Strong Cosmic Censorship} (SCC) conjecture,
this issue regarding the generic nature of solutions of Einstein's
equations which are geodesically incomplete is one of the
outstanding questions of mathematical relativity. Proposed during
the late 1960s by Penrose, the SCC conjecture can be stated as
follows: \emph{Among all sets of constraint-satisfying initial data
whose maximal developments%
\footnote{As proven by Choquet--Bruhat and Geroch \cite{CBG}, for
every initial data set which satisfies the constraints, there is a
unique (up to diffeomorphism) globally hyperbolic spacetime which is
evolved (via Einstein's equations) from that initial data, and which
extends all other globally hyperbolic solutions evolved from the
same data. This spacetime is called the \emph{maximal development}
of that data set. A recent new proof \cite{Sb} of this theorem
avoids the use of Zorn's Lemma.}
are geodesically incomplete, all but a set of measure zero have
unbounded curvature, and cannot be extended $($ across a ``Cauchy
horizon" $)$ as non-globally hyperbolic solutions.} In fact there
\emph{are} many families of globally hyperbolic solutions which are
geodesically incomplete, have bounded curvature, and can be extended
across Cauchy horizons \cite{Mon81}; this is fully consistent with
SCC, since its claim is that \emph{generic} geodesically incomplete
solutions have unbounded curvature and cannot be extended; SCC
allows for \emph{some} solutions to behave otherwise.

As discussed in Chapter 9, Strong Cosmic Censorship has been
carefully formulated and proven for a number of restricted families
of solutions, such as the $T^3$ Gowdy solutions \cite{Ring09}.
Whether or not it holds generally is a open question.

Also open and also one of the central questions of mathematical
relativity is the \emph{Weak Cosmic Censorship} (WCC) conjecture.
Despite its name, the WCC conjecture (also due to Penrose) neither
implies nor is implied by SCC. It can be stated as follows:
\emph{Among all sets of constraint-satisfying initial data sets
whose maximal developments are asymptotically flat and contain a
curvature singularity, in all but a set of measure zero there is an
event horizon which shields the singular region from observation by
asymptotic observers.} In other words, WCC conjectures  that in
solutions of Einstein's equations, gravitational collapse
generically leads to the formation of a black hole. As with SCC,
genericity is a key part of the conjecture: There \emph{are}
gravitationally collapsing solutions with naked singularities, but
WCC conjectures that this does not generically occur. Restricted
families of solutions for which WCC has been proven are discussed in
Chapter 9.

Once it has been determined that a nonlinear hyperbolic PDE system
has solutions which evolve from regular initial data and become
singular in finite time, one of the main questions becomes the
\emph{stability} of solutions. That is, fixing a particular solution
$\Psi$, is it true that the evolution of every solution with
initial data close to that of $\Psi$ must evolve essentially as does
$\Psi$? In addition to its mathematical interest, stability is
important physically, since $\Psi$ is useful in modeling physical
systems only if it is stable in this sense.

For Einstein's equations, stability is not at all obvious. For example,
considering  Minkowski spacetime, does one expect very small
gravitational perturbations of flat data to disperse (stability), or
to concentrate and form a black hole (instability)?

The Christodoulou--Klainerman proof of the stability of Minkowski
spacetime \cite{CK94} is one of the iconic results in mathematical
relativity. Besides resolving an important question concerning
Einstein's theory, this proof showed that many of the difficulties
impeding the analysis of Einstein's equations as a nonlinear PDE
system could be overcome. It showed that coordinate freedom could be
controlled using null foliations, and it showed that the
Bel--Robinson tensor could be used to construct an effective energy
functional for Einstein's theory (see details in Chapter 9). Many of
the ideas developed in this work were used subsequently by
Christodoulou \cite{Chr09} to prove that certain (``pulse") classes
of regular initial data necessarily evolve into spacetimes with
trapped surfaces and therefore very likely into black holes.

Minkowski spacetime is not the only solution of Einstein's equations
which has been shown to be stable. Almost ten years before the
Christodoulou--Klainerman work, Friedrich proved \cite{F86} that the
de Sitter spacetime, which is a solution of the Einstein equations
with positive cosmological constant, is stable. To do this, he
relied on his conformal reformulation of the field equations (based
on the higher--order Bianchi identities) and the conformal
properties of the de Sitter spacetime, thereby avoiding some of the
problems encountered in proving long-time existence for solutions
near Minkowski spacetime. Stability  has also been proven for a
number of solutions which have a steady or accelerating rate of
expansion to the future. As discussed in Chapter 9, Andersson and
Moncrief \cite{AM04} have shown that the Milne
spacetimes%
\footnote{The Milne spacetimes are obtained by topologically
compactifying the constant mean curvature hyperboloids in the future
lightcone of a point in Minkowski spacetime. Together these CMC
hypersurfaces foliate the spacetime, which is flat but is
geodesically incomplete to the past. To the future, the CMC
hypersurfaces have a steady rate of expansion.}
are stable. In this work, a coordinate choice which results in an
elliptic-hyperbolic form for the evolution equations plays a crucial
role. In more recent work, Ringstr\"om \cite{Ring08, Ring09a} has
examined solutions of certain Einstein--scalar--field theories which
are characterized by accelerated rates of expansion and shown that
these solutions are stable. Followup work \cite{Sv, LI13} suggests
that his techniques are robust for expanding solutions.

One of the most challenging open questions in general relativity is
whether the Kerr solutions are stable. This is a mathematically very
rich problem, which has motivated a significant amount of research
(some discussed in Chapter 9). Since the Kerr solutions are believed
by many to accurately model the final state resulting from the
gravitational collapse of large mass stars, it is very important
physically to determine if the Kerr solutions are stable. We note
that general perturbations of Schwarzschild solutions are \emph{not}
expected to evolve to Schwarzschild solutions; rather they are
expected generally to evolve to Kerr solutions.

Stability and cosmic censorship are not the only questions arising
in the study of the long-time behavior of solutions of Einstein's
equations. One would like to explore the long--time, asymptotic
behavior of families of solutions which are not necessarily small
perturbations of known solutions like Minkowski, Kerr, Milne, or
FLRW. While much less is known about how to do this, two very active
areas of research that  are discussed in Chapter 9 could be very
helpful. In part 2, efforts to reduce as much as possible the
regularity (differentiability) needed for well--posedness  are
discussed. Such efforts, which have steadily progressed from
Choquet--Bruhat's original theorem (requiring smooth data)  to the
recent remarkable $L^2$ bounded curvature version
\cite{KRS}, are important for long-time evolution studies because
they establish crucial break--down and continuation criteria for
evolving solutions. That is, if one can prove well--posedness for
data with bounded curvature, then one knows that evolution continues
so long as indeed the curvature remains bounded. Combining results
of this sort with the control of certain energy functionals
(designed to control curvature) would lead to long--time existence.

A scenario for defining and establishing control of energy
functionals is outlined in part 3 of Chapter 9. For solutions
involving negative scalar curvature metrics, one can choose certain
gauges and coordinates, and carry out a reduction%
\footnote{Such a reduction effectively solves some of the
constraints and implements the coordinate and gauge conditions in
such a way that the evolution projects down to a phase space with
fewer dynamical variables. See part 3 of Chapter 9 for details.}
which results in a Hamiltonian formulation with a well-defined,
monotonic Hamiltonian (energy) functional. Combining this reduction
with lessons learned in studying long--time existence for the
Yang--Mills equations \cite{EM82} could lead to very interesting
insights into the dynamics of Einstein's equations.

In mathematical analyses of PDE systems, the emphasis is usually on
understanding the properties and the behavior of large sets of
solutions. For a PDE system used to model physics, one is interested
in individual solutions as well. Numerical simulation of solutions
is crucial for such studies; consequently much effort has gone into
developing numerical relativity since the early 1970s.

Although the techniques used for carrying out numerical simulations
are very different from those used in performing mathematical
analysis, the same basic issues---controlling diffeomorphisms,
choosing gauges and coordinates, distinguishing physical  effects
from coordinate effects,  dealing with a non--fixed background
spacetime---have caused difficulties in both enterprises.
Consequently,  advances in one of these enterprises have often been
very helpful in the other. A key example of this is the
mathematical development of generalized harmonic coordinates as a
tool for proving well--posedness, followed by the crucial role
played by this coordinate choice in carrying out the first stable
numerical simulation of a binary black hole coalescence \cite{Pr05},
in turn followed by the use of generalized harmonic coordinates in
proving the stability of solutions with accelerated expansion
\cite{Ring08}.

Throughout most of its history, from the 1960s to the present,
numerical relativity has been primarily focused on simulating the
coalescence of compact binaries, and determining the consequent
generation of gravitational radiation as well as the properties of
the remaining object. This is largely because of the central role
such collisions are expected to play in generating radiation
observable by LIGO, Virgo and other detectors, and the need to
understand the detailed features of this radiation if it is to be
detected in practice. Despite significant investments in both human
effort and resources, it was a long trail from the earliest efforts
to simulate black hole collisions by Hahn and Lindquist in 1964
\cite{HL64} and the early development work of Smarr and Eppley
during the 1970s,  to the first stable simulations of these
collisions by Pretorius \cite{Pr05} and independently by groups led
by Campanelli \cite{Ca06} and by Centrella \cite{Ce06} in 2005,
culminating in the current systematic production of simulations of a
large parameter space of collisions of binary systems, including
neutron stars as well as black holes. Chapter 7 presents much of the
story of this difficult (and adventurous)  journey.

In addition to its role in simulating sources of gravitational
radiation, numerical relativity has been very useful for other
explorations of general relativity. Perhaps most noteworthy is the
discovery in such simulations by Choptuik \cite{Ch94} of
\emph{critical phenomena} in gravitational collapse. Critical
phenomena are found by considering the evolutions of each of a
one-parameter ($\lambda$) family of initial data sets, with those
solutions evolving from data with large $\lambda$ collapsing to form
black holes, and those solutions evolving from data with small
$\lambda$ dispersing. At the transition value $\lambda_c$ between
data leading to black holes and  data leading to dispersion, the
evolved solution has been found to have special features including
certain types of (continuous or discrete) time symmetry described by
scaling laws. Critical behavior has been observed (via simulations)
to occur for essentially all such choices of one-parameter families
of data (``universality") and has been found to occur for Einstein's
equations coupled to a very wide variety of source fields. Details
are presented in Chapter 7. Interestingly, while the evidence for
critical behavior from numerical simulations is overwhelming, there
has been no mathematical proof of its existence.

Also noteworthy is the significant role which has been played by
numerical simulations in the study of the behavior of the
gravitational field near the singularity in  families of solutions
defined by $T^2$ or $U(1)$ spatial isometry (including the Gowdy
solutions). These simulations, initiated by Berger and Moncrief
\cite{BM93}, indicated the prevalence of ``AVTD" behavior in Gowdy
solutions as well as in polarized $T^2$ symmetric and polarized
$U(1)$ symmetric solutions.  That is, as one evolves toward the
singularity in these solutions (at least in certain chosen
coordinate systems), spatial derivative terms in the evolution
equations appear to be dominated by time derivative terms;
consequently, an observer following a coordinate path sees the
fields evolve more and more like a Kasner solution, with each
observer seeing an independent Kasner. This somewhat peculiar
behavior was first described during the 1960s by Lifshchitz and
Khalatnikov who predicted that it should characterize large classes
of solutions. In later years with Belinskii \cite{BKL} they amended
their prediction to claim that solutions with singularities would
generically exhibit what is now known as ``BKL" or ``Mixmaster"
behavior: each observer would see an endless sequence of Kasner-like
evolutions, punctuated by ``bounces" leading from one Kasner era to
another. While it was widely believed during the 1980s that neither
AVTD nor BKL behavior was likely to be seen, the numerical work by
Berger, Moncrief, Garfinkle, Weaver, Isenberg, and others during the
1990s has supported the contention that AVTD and BKL behavior are
there, but with the added feature of ``spikes". Indeed, it has now
been proven by Ringstr\"om \cite{Ring09b} that $T^3$ Gowdy solutions
\emph{are} generically AVTD; this demonstration is a crucial step in
his proof that Strong Cosmic Censorship holds for these Gowdy
solutions.

Chapter 7 discusses a number of other important numerical relativity
projects which have indicated surprising physical phenomena in
general relativity. Most notable is the apparent instability of
anti-de Sitter spacetime which has been discovered numerically
\cite{Biz14}. This result is very surprising for two reasons: i) it
contrasts with the proven stability of Minkowski spacetime and De
Sitter spacetime, and ii) it has been shown that anti-DeSitter
spacetime is linearly stable.%
\footnote{A solution is linearly stable if the corresponding
linearized Einstein equation operator has negative eigenvalues.}
Also very interesting are the studies of the complex dynamics of
perturbations of black strings in $4+1$ dimensions \cite{LP10}. In
both of these cases, mathematical questions have motivated the
studies, and the numerical work has provided very strong indications
of unambiguous answers to these questions.

One area not discussed in these Chapters in which the partnership of
numerical and mathematical work has been very useful is in the study
of the strong field regime during the formation  and coalescence of
black holes. Mathematically, the definition of black holes has
traditionally been associated with event horizons, the future
boundaries of the causal past of future null infinity. They separate
the spacetime region from which light can escape to infinity from
those regions from which light can \emph{never} do so.
Unfortunately, an immediate consequence is that the notion of an
event horizon is not only extremely global but it is also
teleological; one has to know the entire spacetime before one can
locate it. In particular, an event horizon can grow in regions which
have weak---or even zero---spacetime curvature, \emph{in
anticipation} of a future gravitational collapse. Therefore in the
study of gravitational collapse or dynamics of compact objects, one
cannot locate event horizons \emph{during} evolution. In practice,
numerical simulations have used ``apparent horizons" on the chosen
family of Cauchy hypersurfaces to indicate incipient black holes
because neither of the two light fronts emanating from such surfaces
is expanding; light is trapped at least instantaneously. However,
this notion is tied to the choice of Cauchy hypersurfaces. To get
around this issue, one can consider spacetime world tubes of
apparent horizons, known as \emph{dynamical horizons} if they
evolve, and \emph{isolated horizons} if they do not
\cite{hayward,aaetal,akrev,gjrev}. As black hole markers, these
structures have the advantage that they are determined by the local
spacetime geometry in their immediate vicinity; they do not require
knowledge of the full spacetime and in particular they do not exist
in a flat spacetime. Dynamical horizons and isolated horizons have
been studied both analytically and numerically. From these studies,
tools have been developed to extract physical information, such as
the mass, angular momentum, and (source) multipole moments of black
holes, even in dynamical regions, in a coordinate invariant fashion.
These structures have also made it possible to meaningfully compare
results of simulations that use different initial data, coordinates
and foliations in the strong field region. Finally, they provide
analytical tools to probe \emph{how} back holes approach their
final, universal equilibrium state, represented by the geometry of
the Kerr horizon, and if the process has universal features
\cite{ac,ro}. While these developments were inspired by simulations,
they have also had interesting applications on the analytical side,
including significant generalizations of the laws of black hole
mechanics \cite{bch,leshouches} to fully dynamical situations
\cite{akrev}, black hole evaporation \cite{apr} and calculations of
horizon entropy using quantum geometry, as summarized in Chapter 11.

One very elusive issue for which dynamical horizons could be helpful
is to find  a comprehensive definition of black holes which makes
sense in dynamical spacetimes, as well as in spacetimes which are
not asymptotically flat \cite{core}. One problem with using
dynamical horizons, however, is that they are generally not unique
\cite{ag}; consequently they are not good candidates for canonically
locating the surface of a black hole.

Finally, we  note one other very elusive physical issue which calls
for both numerical and mathematical studies, in partnership,
concerning gravitational waves.  As explained in Part II, although
Einstein analyzed gravitational waves in the linearized
approximation and derived the quadrupole formula soon after his
discovery of the field equations, there was considerable confusion
over the next three decades regarding the physical reality of
gravitational waves because it was difficult to disentangle physical
effects from coordinate artifacts. The situation was clarified only
in the sixties when the notion of \emph{Bondi news} was shown to
provide a coordinate invariant characterization of gravitational
radiation \cite{gr3}. That framework continues to provide the
conceptual foundations of the gravitational wave theory discussed in
Chapter 6. However, it assumes a zero cosmological constant;
$\Lambda=0$. This is a problem, because cosmological observations
have established that  $\Lambda >0$ is to be preferred and, surprisingly, we do not
yet have a satisfactory analog of the Bondi news, or viable
expressions of energy, momentum and angular momentum carried by
gravitational waves in full, non-linear general relativity for
$\Lambda>0$. Indeed, even the asymptotic symmetries that are needed
to introduce these notions do not extend to the $\Lambda >0$ case.
In cosmological applications, it has been sufficient just to use
linearized gravitational waves. Similarly, for isolated systems such
as compact binaries or collapsing stars, working with $\Lambda=0$
should be an excellent approximation in most calculations since the
observed value of $\Lambda$ is so small. However, as of now, we do
not have any theory for $\Lambda >0$, whose predictions are to be
approximated by the $\Lambda=0$ calculations. Only when we have this
theory and a detailed understanding of how $\Lambda$ affects
physical quantities can we be confident that the current $\Lambda=0$
calculations do approximate reality sufficiently well. Moreover,
there may well be subtle effects---similar to the Christodoulou
memory effect in the $\Lambda=0$ case---that have eluded us and
which will be uncovered only when the $\Lambda >0$ case is well
understood.

For a number of years following its pristine birth one hundred years
ago, general relativity was all too often admired and adulated
rather than tested and put to work. Just as this has now changed in
the experimental and observational realm, as reported in Parts I and
II, we see here in Part III that this has dramatically changed in
the mathematical and numerical realm as well. Einstein's theory is
now recognized as a fresh source of interesting and challenging
mathematical problems, and the successful solution of some of these
problems has attracted increasing numbers of mathematicians with
innovative skills. In turn, with more powerful analytical tools
available, relativists have been able to explore the physical
implications of general relativity more effectively. The intriguing
mix of beautiful mathematics and profound physics, always an
enticing feature of general relativity, is now a central reason for
expecting rapid progress both in its use as an effective tool for
studying gravitational physics, and its role as a challenging field
of geometric analysis.

\section{Part IV: Beyond Einstein}
\label{s5}

The remarkable advances summarized in the first three parts of this
volume refer almost entirely to the well-established realm of
classical general relativity (GR). However, Einstein \cite{ae} was
quite aware of the limitations of his theory. In the context of
cosmology he wrote, as early as in 1945,
\begin{quote}
{\sl ``One may not assume the validity of field equations at very
high density of field and matter and one may not conclude that the
beginning of the expansion should be a singularity in the
mathematical sense."}
\end{quote}
By now, we know that classical physics cannot always be trusted even
in the astronomical world because quantum phenomena are not limited
just to tiny, microscopic systems. For example, neutron stars owe
their very existence to a quintessentially quantum effect: the Fermi
degeneracy pressure. At the nuclear density of $\, \sim 10^{15}\,
{\rm g/cm^{3}}$ encountered in neutron stars, this pressure becomes
strong enough to counterbalance the mighty gravitational pull and
halt the collapse. The Planck density is some \emph{eighty} orders
of magnitude higher! Astonishing as the reach of GR is, it cannot be
stretched into the Planck regime; here one needs a grander theory
that unifies the principles underlying both general relativity and
quantum physics.\\

\textbf{Early developments}\\

Serious attempts at constructing such a theory date back to the
1930s with papers on the quantization of the linearized
gravitational field by Rosenfeld \cite{rosenfeld} and Bronstein
\cite{bron1}. Bronstein's papers are particularly prescient in that
he gave a formulation in terms of the electric and magnetic parts of
the Weyl tensor and his equations have been periodically
rediscovered all the way to 2002 \cite {bron2}! Analysis of
interactions between gravitons began only in the 1960s when Feynman
extended his calculational tools from QED to general relativity
\cite{rpf}. Soon after, DeWitt completed this analysis by
systematically formulating the Feynman rules for calculating the
scattering amplitudes among gravitons and between gravitons and
matter quanta. He showed that the theory is unitary order by order
in perturbation theory (for summary, see, e.g., \cite{bsd}). In
1974, 't Hooft and Veltman \cite{thv} used elegant symmetry
arguments to show that pure general relativity is renormalizable to
1 loop but they also found that this feature is destroyed when
gravity is coupled to even a single scalar field. For pure gravity,
there was a potential divergence at two loops because of a counter
term that is cubic in the Riemann tensor. However there was no
general argument to say that its coefficient is necessarily
non-zero. A heroic calculation by Goroff and Sagnotti \cite{gs}
settled this issue by showing that the coefficient is $(209/2880
(4\pi)^2)$! Thus in perturbation theory off Minkowski space, pure
gravity fails to be renormalizable at 2 loops, and when coupled to a
scalar field, already at 1-loop.

The question then arose whether one should modify Einstein gravity
at short distances and/or add astutely chosen matter which would
improve its ultra-violet behavior. The first avenue led to higher
derivative theories. Stelle, Tomboulis and others showed that such a
theory can be not only renormalizable but asymptotically free
\cite{asymfree}. But it soon turned out that the theory fails to be
unitary and its Hamiltonian is unbounded below. The discovery of
supersymmetry, discussed in Chapter 12, suggested another avenue:
with a suitable combination of fermions and bosons, perturbative
infinities in the bosonic sector could be canceled by those in the
fermionic sector, improving the ultraviolet behavior. This hope was
shown to be realized to 2 loops by Deser et al and Grisaru et al \cite{sugra1}. However, by the late 1980s a consensus emerged that
all supergravity theories would diverge by 3 loops and are therefore
not viable (see, e.g., \cite{howe-stelle}). \\

A series of parallel developments was sparked in the canonical
approach by Dirac's analysis of constrained Hamiltonian systems. In
the 1960s, this framework was applied to general relativity by
Dirac, Bergmann, Arnowitt, Deser, Misner and others
\cite{adm,komar,pbak,agrev,kk1}. The basic canonical variable was
the 3-metric on a spatial slice and, as discussed in Chapters 8-10,
general relativity could be interpreted as a dynamical theory of
3-geometries. Wheeler therefore baptized it \emph{geometrodynamics}
\cite{jw1,jw2}. Wheeler also launched an ambitious program in which
the internal quantum numbers of elementary particles were to arise
from non-trivial, microscopic topological configurations and
particle physics was to be recast as \emph{`chemistry of geometry'}.
This led to interesting discoveries at the interface of topology and
general relativity but the approach did not have notable success
with the particle physics phenomenology.

A distinguishing feature of the canonical approach is that in
contrast to perturbative treatments it does not split the metric
into a kinematic background and a dynamical fluctuation. As a
result, a number of conceptual problems were brought to the
forefront which revealed the deep structural differences between
general relativity and more familiar field theories in Minkowski
spacetime. By now there is a near universal appreciation of the
importance of background independence and of the necessity of facing
the ensuing complications. However, this very feature made it
difficult to use the standard techniques from QED to face the
mathematical difficulties associated with the infinite number of
degrees of freedom of the gravitational field. Consequently, most of
the work in full quantum geometrodynamics remained rather formal.
Detailed calculations could be carried out in the context of quantum
cosmology where one freezes all but a finite number of degrees of
freedom. Initially there was hope that quantum effects would tame
the cosmological singularities of general relativity. However, this
hope did not materialize; even in the simplest models the big bang
could not be softened without additional `external' inputs into the
theory. The program also faced a sociological limitation in that the
ideas that had been so successful in QED played no role: in a
non-perturbative, background independent approach, it is hard to see
gravitons, calculate scattering matrices and use virtual processes
to obtain radiative corrections. To use a well-known phrase \cite
{weinberg}, the emphasis on geometry in the canonical program
``drove a wedge between general relativity and the theory of
elementary particles." Therefore, after an initial burst of
activity, the quantum
geometrodynamics program became rather stagnant.\\

A third avenue was opened in the mid 1950s: explorations of the
effects of a \emph{classical} gravitational field on quantum matter
fields. Early work by Parker explored quantum fields in FLRW spacetimes \cite{parker}. As
recent successes of inflationary scenarios discussed in Chapters 4
and 11 show, this choice was prescient. Indeed, this is the arena
where we are most likely to first see the interface of gravity and
quantum physics observationally. But this general area did not draw
much attention until Hawking's seminal discovery in 1974 that
quantum field theory (QFT) on a black hole background predicts that
black holes emit quantum radiation and resemble black bodies when
seen from infinity. Not only did the entire area of QFT in curved
spacetime experience an explosion of activity but this discovery
has served as a focal point for a great deal of research in all
areas of quantum gravity over the last four
decades.\\

\textbf{Current status}\\

Ideas developed in QFT in curved spacetimes have had a number of
fascinating applications, ranging from the study of diverse aspects
of the Casimir effect \cite{casimir} to the feasibility studies of
creating time machines by exploiting the violations of local energy
conditions that are allowed in QFT \cite{timemachines}. Advances on
the more fundamental side are discussed in Chapter 10. In general
curved spacetimes, we do not have the Poincar\'e group to decompose
fields into positive and negative frequency parts and select a
canonical vacuum. Since there is no natural choice of a (Fock)
representation of the canonical commutation relations, one is led to
the \emph{more general} setting of algebraic QFT. As a consequence,
recent advances in QFT in curved spacetimes have brought to the
forefront the essential conceptual ingredients of QFT that are often
masked by the extraneous structure that happens to be available in
Minkowski spacetime. As emphasized in Chapter 10, these
developments have also shown that, beyond the special context of
static spacetimes, techniques from Euclidean QFT cannot be carried
over to calculate quantities of direct physical interest in the
Lorentzian theory. This is an important message also for approaches
to full quantum gravity although, unfortunately, it is often
overlooked. On the physical side, quantum effects on curved
spacetimes lead to unforeseen phenomena such as evaporation of
black holes and emergence of the large scale structure of the
universe from pure quantum fluctuations in the very early universe.
Finally, advances in describing interactions have now elevated QFT
in curved spacetimes to the same mathematical level as
that enjoyed by rigorous QFT in Minkowski space.\\

In quantum gravity proper, while both the perturbative and the
canonical approaches reached an impasse by the early 1980s, they
provided seeds for most of the subsequent developments. Although GR
is perturbatively non-renormalizable, an effective field theory was
developed systematically \cite{effective} and has had remarkable
successes in the low energy regime, e.g., in the treatment of
dynamics of compact binaries in classical GR \cite{rothstein} and in
the computation of the leading corrections to the Newtonian
potential. Therefore the problem of finding a viable quantum gravity
theory can be rephrased as that of obtaining an appropriate
completion of this theory in which the outstanding conceptual issues--such as the fate of the classical singularities of GR, the
statistical mechanical accounting of black hole entropy, and the
final stages of the black hole evaporation---can be analyzed
systematically.

This quest has been undertaken in a number of directions. Each
program adopts a different point of departure, treating certain
aspects of the problem as more fundamental, and hoping that the
remaining aspects can be handled successfully once there is a
resolution of the key difficulties. Chapter 11 focuses on approaches
which emphasize the dynamical nature of spacetime geometry and 
non-perturbative methods. The point of departure is GR. However, the
fundamental degrees of freedom and the short-scale dynamics in the
final quantum theory are quite different from those of GR and
classical spacetimes and gravitons emerge only in a suitable limit.

The first of these programs is Asymptotic Safety, whose goal is to
provide a specific \emph{ultraviolet completion} of the effective
field theory using Weinberg's generalized notion of renormalization.
On the analytical side, non-trivial fixed points of the
renormalization group flow have been obtained in 4 spacetime
dimensions, even after allowing for 9 different gravitational
couplings in addition to the coupling of gravity to matter fields.
These results strongly suggest that much of the intuition derived
from Goroff and Sagnotti's early results are tied to perturbation
theory around Minkowski space. Consistent ultraviolet completions of
the standard effective theory may well exist \emph{without} having
to invoke supersymmetry, higher dimensions or extended objects.
Similarly, on the computational side it was widely believed that the
renormalization group flows would inevitably lead to a `crumpled
phase' in which macroscopic spacetimes such as the one around us
will not emerge in the infrared. Recent progress in Causal Dynamical
Triangulations has shown that the `crumpled phases' are not
inevitable; theory does allow smooth, macroscopic spacetimes with
small quantum fluctuations.

The second program discussed in Chapter 11 is Loop Quantum Gravity
(LQG) which grew out of the canonical approach. Consequently,
manifest background independence is at the forefront and the theory
is again nonperturbative. However, emphasis is shifted from metrics
to connections; the `wedge' between general relativity and gauge
theories governing other fundamental forces is removed. Indeed, the
basic notions are taken directly from the Yang-Mills theory but
without reference to a background spacetime metric. Rather
surprisingly, this strategy leads to a unique kinematic setup in
which one can overcome the mathematical obstacles faced in
geometrodynamics and handle the infinite number of degrees of
freedom rigorously. The Hilbert space is spanned by \emph{spin
networks}. Consequently, the fundamental excitations of geometry/
gravity are not gravitons which represent quantum fluctuations on a
given background geometry. Rather, they are polymer-like threads
that can be woven \emph{to create the geometry itself.} Geometry is
quantum mechanical in the most direct sense: geometric operators
have discrete eigenvalues. This discreteness has a deep influence on
dynamics. In covariant LQG---called the Spinfoam framework---it
offers relief from ultraviolet infinities by banishing degrees of
freedom at scales shorter than the Planck length. A priori, there
could be infrared infinities but, surprisingly, if there is a
\emph{positive} cosmological constant, they can be removed by a
natural quantum deformation of the local Lorentz group. LQG has had
notable success both in the cosmological and black hole sectors of
GR. In both cases, the underlying quantum geometry plays a key role.
In particular, in the cosmological sector it naturally tames all
strong curvature singularities even when matter satisfies all the
energy conditions. Furthermore, the framework has extended the reach
of observational cosmology all the way to the Planck regime through
the development of QFT on \emph {quantum} cosmological spacetimes.
A similar extension is now being used to study Hawking
radiation on \emph{quantum} black hole spacetimes.\\

Chapter 12 summarizes advances based on supersymmetry, extended
objects, higher dimensions and holography. Chronologically, at first
the emphasis in the framework was on perturbation theory in
Minkowski spacetime. The viewpoint was that the ultraviolet
divergences of quantum gravity were signals that point particles and
local quantum fields are over-idealized notions that become
untenable at very high energies. Fundamental objects are strings,
and interactions between them are just the simple processes of
joining and splitting which naturally avoid the ultraviolet
infinities of local QFT. Particles are merely excitations of
strings. This immediately leads to an infinite tower of particles
and fields but most of these excitations are so massive that they
become relevant only at extremely high energies. Theoretical
consistency implies that spacetime has to be 26-dimensional without
supersymmetry and 10-dimensional with supersymmetry. At first it was
believed that the extra dimensions could be compact and microscopic,
and that there would be severe constraints on permissible
compactifications as well as on the matter content and allowed
interactions to make the theory unique. Therefore it was often
heralded to be the `theory of everything'.

Further research showed that these expectations were overly
optimistic, but at the same time it revealed richer structures: The
theory turned out \emph{not} to be unique but various consistent
theories are related by certain dualities. It is now believed that
there \emph{is} probably a single theory and the known consistent
theories represent its `corners'. However, as of now its structure
remains opaque; indeed, one does not even know what the fundamental
principles behind it should be. This is reflected in its very name,
\emph{`M-theory'}, where M is often said to stand for `Mystery'.
Also, most of the work to date has been carried out in the framework
of first quantization and the possible role of a second quantized
string field theory remains unclear. But it \emph{is} clear that the
theory contains not only strings but also higher dimensional,
extended, non-perturbative configurations, called $p$ branes. This
enlargement of the theory opened a gate to incorporate extremal (and
certain near-extremal) black holes in the string paradigm. Because
of the `non-renormalization' theorems, it is possible to calculate
the number of certain string states at low energy and then correctly
deduce the number of microstates of these black holes.

These considerations in turn suggested the possibility of a
holographic picture in which string theory on asymptotically anti-de
Sitter backgrounds could be regarded as being dual to certain
conformal field theories on the boundary of these spacetimes. As
discussed in Chapter 12, this AdS/CFT conjecture has led to a wealth
of insights on the unity of the mathematical structures that
underlie completely distinct physical systems such as quark-gluon
plasmas and condensed matter systems. In addition, throughout its
evolution, string theory has had unforeseen applications to diverse
areas of mathematics. Given that these ideas extend the reach of
gravitational physics to completely new territories in ways that
were not even imagined before, it is fitting that the volume
concludes with this Chapter.\\

Finally, it is important to emphasize that the content of Part IV
does not do full justice to the field because space limitation did
not allow us to include several promising advances. First, even in
the areas covered in Chapters 10-12, in the spirit of this volume,
authors focused on a few topics on which most significant advances
have occurred over the last three decades or so. The second and more
important omission is that several approaches to quantum gravity had
to be left out entirely. In the first three Parts of this volume,
which are based on well-established ideas and careful observations,
the choice of what to include was easier to make. In the case of
quantum gravity, the subjective element is much more pronounced
simply because one is now forced to leave the safety net of ideas
that are firmly grounded in GR. Promising directions that were left
out include: \textit{i)} Asymptotic quantization, which brought out
the interplay between the Bondi-Metzner-Sachs group and infrared
issues in full quantum gravity \cite{asymquan}; \textit{ii)} Twisor 
theory based programs for calculating scattering amplitudes from past to future null infinity, that are now drawing a great
deal of attention \cite{scattering}. Twistor theory itself has
provided a powerful bridge between the theory of partial
differential equations and algebraic geometry which extends to
(self-dual) Einstein's equations through Penrose's non-linear
graviton construction \cite{rpwr}; \textit{iii)} The Regge calculus
approach, which parallels lattice QCD, but uses dynamical simplicial
decompositions rather than lattices defined in a background geometry
\cite{hamber}; \textit{iv)} Ho\v{r}ava-Lifshitz gravity, which
sacrifices manifest local Lorentz invariance to achieve better
ultraviolet behavior, hoping to recover it in the infrared limit
\cite{horava}; \textit{v)} Causal sets, in which one postulates that
at a fundamental level one only has a discrete set of points with
causal relations between them \cite{causets}; and \textit{vi)} The
Vassiliev higher spin theories, in which an infinite tower of
\emph{massless} higher spin fields are incorporated in a consistent
manner \cite {vassiliev}. One or more of these ideas may well lead
to an ultraviolet completion of GR with desired features.

In this regard, it is instructive to look back at the events that
celebrated the centennial of Einstein's birth some 35 years ago. The
Princeton conference at the Institute of Advanced Study had two
talks on quantum gravity; both on supergravity \cite{wolf}. The
Cambridge University Press volume \cite{hi} also had two Chapters,
one that introduced the asymptotic safety program and the other that
put all its emphasis on Euclidean quantum gravity. A year later, in
his Lucasian Chair inaugural address, Hawking \cite{swh1} suggested
that the end of theoretical physics was in sight because $N=8$
supergravity was likely to be the final theory. The field has
evolved rather differently! Fascinating as the advances over the
last three decades have been, the
reader would do well to keep this historic perspective in mind.\\

\textbf{Elephants in the room}\\

There is no question that today we understand the interface of
gravity with quantum physics much better than we did in the mid
1980s when the approaches discussed in Chapters 11 and 12 first rose
to prominence. Several unworkable ideas have been weeded out and, as
our summary indicates, concrete advances have provided us with novel
insights. It is therefore fitting that the review articles and
status reports tend to be  upbeat, exuding confidence. But it is
also clear that the end of quantum gravity is not yet in sight. We,
the Editors, would be remiss if we do not venture to say why.

The leading approaches use diverse points of departure and have
strikingly different perspectives as to what is most fundamental and
what can be revisited later. The mathematical techniques they use
also vary significantly. Given the difficulty of the task, this
diversity is of course both healthy and essential. As our summary of
early developments shows, this diversity has been a hallmark of this
field for over 50 years. Today, the practitioners are even more
passionate about the choices their approach makes. However, as a
result, whereas other areas of gravitational science have witnessed
a convergence of ideas and coming together of previously distinct
communities, in quantum gravity the communities have drifted further
apart. As active communications become less frequent, slowly but
steadily the tendency to ignore the elephants in one's room
increases. At the same time, it seems more and more natural to think
of other viewpoints as untenable. To make this point more concrete,
we will provide a few illustrative examples from the main programs
discussed in this volume.

In the approaches based on unification, ideas that lead to
mathematically rich structures play a dominant role. Consequently,
as we discussed above, these approaches have led to unforeseen
insights into the mathematical unity in the description of diverse
physical systems. However, this success also seems to have fueled a
tendency to ignore the issue of whether the central ideas behind
these approaches are realized in our physical universe. In
particular, there is little hesitation in building a quantum theory
of gravity by demanding that it have supersymmetry, higher
dimensions, infinite towers of particles and fields and a negative
cosmological constant \emph{at its very foundation}. To researchers
outside quantum gravity the strategy seems surprisingly indifferent
to the current state of observations, rather akin to searching for the
key under a lamppost irrespective of where it was actually lost. It
is true that some of these ideas are motivated by the Kaluza-Klein
theory where extra dimensions are meant to be microscopic and curled
up, remaining invisible until we reach energies near the Planck
scale. But in detailed explorations this primary restriction is
often set aside. For example, it is rarely imposed while studying
higher dimensional solutions in these frameworks. More importantly,
in the most commonly used versions of the AdS/CFT conjecture,
symmetry requirements force the extra dimensions to be very large;
the radius of the compactified internal spheres is \emph {the
cosmological radius}! If we lived in such a universe, the internal
dimensions should be as readily observable as the `normal' 4
spacetime dimensions. Taken together, these assumptions push one to
a paradigm whose relevance to the physical issues of quantum gravity
in the actual universe we inhabit becomes increasingly obscure.

Approaches developed primarily by researchers from the GR community
make a serious attempt to base their foundations only on the well
established principles of GR and QFT. They tend to take the
gravity/geometry duality seriously and aim to understand the quantum
nature of geometry first, postponing the issue of coupling to matter
to a second stage. This strategy has had notable success in the
asymptotic safety program. However, in this approach exploration of
the quintessentially quantum gravity issues---such as the origin of
black hole entropy and the fate of the most vexing singularities of
GR---is still at a preliminary stage. Furthermore, whereas at a
fundamental level the program refers to renormalization group flows
in the infinite dimensional space of permissible theories, in
practice it seems unlikely that one would be able to go beyond
finite truncations in any foreseeable future. Therefore the question
of whether the program truly provides the promised ultraviolet
completion is likely to remain open for a long time.

LQG has also advanced by making truncations. However, as in the
concrete calculations of QED or QCD, truncations refer to physical
problems of interest, such as the Planck scale physics of the very
early universe or quantum properties of black holes. In these
truncated sectors, the theory has had notable success. But, as
discussed in Chapter 11, the issue of dynamics in full LQG is still
far from being settled. Spinfoams provide a natural framework to
explore it and certain Spinfoam models have had notable success. But
a number of fundamental issues remain: Does the natural expansion
used to calculate `transition amplitudes' converge? Is a continuum
limit needed and, if so, how exactly is one to take it? Do the
proposals to incorporate matter in Spinfoams work in detail? An
equally important open issue is to make direct contact with low
energy physics. Since one begins with quantum geometry in the Planck
regime and then descends to the low energy world, this problem is
highly non-trivial. There are numerous preliminary, encouraging
results, such as the calculation of the graviton propagator starting
from a fully non-perturbative and background independent setting.
But the relation to the effective field theory beyond the leading
approximation remains unclear. Until there is a solid bridge linking
non-perturbative dynamics to the well-developed effective theory in
detail, physical viability of the approach will remain uncertain.
\\

\textbf{Epilogue}\\

While writing a review article on special relativity in 1907,
Einstein realized that Newtonian gravity is incompatible with
special relativity and set himself the task of resolving this
conflict by creating a grander synthesis from which the two theories
would emerge as limiting cases. Just eight years later, he arrived
at the finished solution and for the last century, physical
scientists from a broad array of disciplines have been happily
engaged in investigating its content! Work on unifying GR with
quantum physics, on the other hand, has seen many twists and turns;
periods of euphoria followed by despair at conceptual impasses, or
Nature's stubborn refusal to use structures that seem compelling to
the practitioners. Einstein's spectacular success is in striking
contrast with the time and effort that has been devoted to this
endeavor.

But it is important to note that progress in physical theories has
more often mimicked the development of quantum theory rather than
general relativity. More than a century has passed since Planck's
discovery that launched the quantum. Yet, the theory is incomplete.
We do not have a satisfactory grasp of the foundational issues,
often called the `measurement problem'. Nor do we have a single
example of a mathematically complete, interacting QFT in 4
dimensional Minkowski spacetime. A far cry from general relativity
that Einstein offered us in 1915! Yet, no one would deny that
quantum theory has been extremely successful; indeed, its scientific
reach vastly exceeds that of general relativity.

Thus, while it is tempting to wait for another masterly stroke like
Einstein's to deliver us a finished quantum gravity theory, it would
be more fruitful to draw lessons from quantum physics. There,
progress occurred by focusing not on the `final' theory that solves
all problems in one fell swoop, but on concrete physical problems
where quantum effects were important. Experience to date indicates
that the same will continue to be true for quantum gravity in the
foreseeable future.

What could hasten progress along this path? So far individual
programs have been driven by internal criteria. More significant
advances could occur by critically examining common elements they
share as well as tensions between the ideas that lie at their
foundations.

For example, although these programs start with very different
viewpoints and assumptions, they feature a curious dimensional
reduction at the Planck length \cite{carlip}. In LQG, the
fundamental excitations of quantum geometry have 1 spatial dimension
whence Spinfoams are 2-complexes. In Asymptotic Safety one has a
running spectral (spacetime) dimension which (equals 4 in the
infrared but) approaches  2 in the ultraviolet. In perturbative
string theory, point particles are replaced by 1 (spatial)
dimensional strings, propagating on a classical spacetime, say,
Minkowski space. In the linearized approximation off Minkowski
space, the LQG excitations have the same massless particle content
as in bosonic string theory---a dilaton, an antisymmetric tensor
and a spin-2 excitation---to begin with, and the graviton is
extracted by imposing linearized Einstein constraints. Is there
perhaps a deep reason why qualitatively similar 2-dimensional
structures arise even when the starting points are so different?

The next set of issues is related to ultraviolet finiteness. Since
the late 1980s there has been a strong belief in the string theory
community that local interactions between point particles \`a la QFT
would inevitably lead to ultraviolet divergences and the theory is
rendered perturbatively finite by using strings instead. There has
been considerable research on this issue especially in recent years.
However, as the number of loops grow, there is an increasing number
of ambiguities (because of the super-moduli measure needed) in the
calculation and therefore some experts continue to believe that the
issue of order by order perturbative finiteness is still open in
string theory \cite{nicolai}. On another front, recent work on
supergravity, described in Chapter 12, has shown that the N=8
supergravity in 4 spacetime dimensions is finite to 4 loops,
contrary to the near unanimous expectations in the 1980s
\cite{howe-stelle}. Some experts in supergravity have suggested that
there could well be symmetries that have remained hidden so far that
could make supergravity finite to all orders. Support for the
general idea of hidden symmetries comes also from Vassiliev's higher
spin theories. Overall, there is a small but growing community that
believes that what is fundamental for finiteness is a sufficiently
large and subtle symmetry group rather than extended objects. Thus,
there is a healthy tension concerning the issue of what drives
finiteness even at the level of perturbation theory.

The tension continues beyond perturbations. In LQG, background
independence leads to a rather sophisticated quantum geometry with
the property that there are no degrees of freedom below the Planck
length, ensuring ultraviolet finiteness. In string theory, the sum
in the perturbation expansion diverges \cite{gross-periwal}
requiring a non-perturbative treatment. As we saw, a highly
successful candidate is available in the sector of the theory with a
negative cosmological constant: the AdS/CFT conjecture, where string
theory in the bulk is equivalent to a finite field theory on the boundary.%
\footnote{It is interesting, indeed, that the best non-perturbative
definition of string theory takes us back to a local QFT, albeit on
the boundary of spacetime.}
Note that this AdS/CFT correspondence also provides a background
independent definition of string theory in the asymptotically AdS
sector. But whereas the mechanism taming the ultraviolet regime is
provided by a specific quantum Riemannian geometry in LQG, in string
theory it is provided by holography. Is there nonetheless a deep
connection between them? If so, it may be helpful in, e.g., the
analysis of the physical cosmological singularities in string
theory.

The next example pertains to the gravity/gauge theory duality. As we
noted above, LQG starts by reformulating GR in terms of Yang-Mills
type variables and making heavy use of gauge theory notions such as
holonomies, Wilson loops and quantized fluxes. The recent finiteness
results in supergravity rely heavily on the relation between
structures that arise in the N=4 super Yang-Mills theory and N=8
supergravity in 4 dimensions. In the AdS/CFT correspondence this
interplay is also at the forefront; for example, string theory in a
5 dimensional asymptotically AdS background is dual to this super
Yang-Mills theory (on the boundary). Is this interplay between
gravity and gauge theories a beacon guiding us to a rapprochement of
various approaches? Could a deeper understanding of this duality
lead to a new principle which has eluded us because we have examined
the issue only piecemeal, from the perspective of only one approach
at a time?

The last example is provided by the analyses of the statistical
mechanical origin of black hole entropy. In string theory
calculations one generally considers strings with end points of
branes that carry gauge fields. In LQG, the horizon `membrane' is
pierced by the polymer excitations of the bulk quantum geometry and
intrinsic geometry of quantum horizons is described by a gauge
theory. Qualitatively, the two pictures appear to be similar. Yet,
the detailed analyses are very different and their strengths and
limitations are complementary. The string theory  calculations refer
to certain extremal and near extremal black holes, while in LQG they
are based on the notion of isolated horizons and therefore include
all black hole and cosmological horizons. But whereas the relation
to semi-classical calculations of entropy is well understood in
string theory, the issue remains open in LQG because what one counts
is the number of microstates of the quantum horizon geometry which
are conceptually quite distinct from the quantum corrections to the
Euclidean, classical action. Given the qualitative similarity of
concepts underlying these calculations, can one perhaps relate them
in detail? Such a  bridge would enable one to export the strengths
of each of these calculations to overcome the corresponding
limitation of the other. Thus, a better understanding of common
elements, differences and relations between different approaches
could well suggest a paradigm that combines deep ideas from various
approaches.

Over the past three decades, such rapprochements were hindered by
periodic bouts of irrational exuberance which led individual
communities to be certain that theirs was the only viable path and,
by implication, nothing else was worth paying attention to. To
paraphrase the biologist Fran\c{c}ois Jacob, for sustained progress
in any area of science, it is important that the practitioners be
aware of the limits of their science and thus their knowledge.
Otherwise it is easy to mix what one believes and what one knows to
create a misplaced sense of certitude. Ongoing dialogues across
various approaches and careful examinations of the common elements
and differences between them would go a long way to avoid this trap.

\section*{Acknowledgments}

We have profited from discussions with a large number of colleagues,
and correspondence with the authors of the 12 Chapters in this
volume. This work was supported in part by the NSF grants
PHY-1205388, PHY-0968612, PHY-1306441 and FRG-DMS-1263431, and the Eberly research funds of Penn state.


\begin{thebibliography}{99}
\expandafter\ifx\csname
natexlab\endcsname\relax\def\natexlab#1{#1}\fi
\expandafter\ifx\csname selectlanguage\endcsname\relax
  \def\selectlanguage#1{\relax}\fi


\bibitem{Edd18} A. Eddington, Report on the relativity theory of
    gravitation. (Fleetway Press, London, 1918); Report to the
    Physical Society of London, 1Reprinted 1920.

\bibitem{Har79} G.~Harvey, Observatory \textbf{99}, 195–8 (1979).

\bibitem{HawEll73} S.~W.~Hawking and G.~F.~R.~Ellis, \textit{The
    large scale structure of space-time} (Cambridge University Press, Cambridge,
    1973).

\bibitem{Bon60} H.~Bondi, \textit{Cosmology}  (Cambridge
    University Press, Cambridge, 1960)

\bibitem{Tau51} A.~Taub, Ann. Math. \textbf{53}, 472 (1951);
    Reprinted,
    with editorial introduction by M.A.H. MacCallum, in Gen. Rel.
    Grav. \textbf{36}, 2689-2719 (2004).

\bibitem{Pet54} A.~Petrov, Scientific Proceedings of Kazan State
    University (named after V.I. Ulyanov-Lenin), Jubilee (1804-1954)
    Collection, 114, 55–69 (1954). Translation by J. Jezierski and
    M.A.H. MacCallum, with introduction by M.A.H. MacCallum, Gen.
    Rel. Grav. \textbf{32}, 1661-1685 (2000).

\bibitem{SteKraMac03} H.~Stephani, et al. \textit{Exact solutions of
    Einstein's field equations}, 2nd edition, (Cambridge University Press, Cambridge 2003); Corrected Paperback edition, (2009).

\bibitem{GriPod09} J.~Griffiths and J~Podolsky, \textit{Exact
    space-times in
    Einstein's general relativity}, Cambridge Monographs in
    Mathematical Physics, (Cambridge University Press, Cambridge,
    2009).

\bibitem{Bru91} V.~Brumberg, \textit{Essential relativistic
    celestial
    mechanics} (Adam Hilger, Bristol 1991).

\bibitem{Lyn69} D.~Lynden-Bell, Nature \textbf{22}, 690 – 694
    (1969).

\bibitem{BonPirRob59} H.~Bondi, F.~Pirani, and I.~Robinson, Proc.
    Roy. Soc. Lond. A\textbf{251}, 519 (1959).

\bibitem{Hur08} K.~Hurley, et al. [LIGO collaboration], Astrophys.
    J. \textbf{681}, 1419–1428 (2008).

\bibitem{Gut81} A.~H.~Guth, Phys. Rev. D\textbf{23}, 347–356 (1981).

\bibitem{Har84} M.~Harwit, \textit{Cosmic Discovery: The Search,
    Scope and Heritage of Astronomy} (The MIT Press, Cambridge 1984).

\bibitem{KruAgo14} E.~Kruse and E~Agol, Science \textbf{344}, 275
    (2014).

\bibitem{ColEll97} P.~Coles and G.~F.~R.~Ellis,  \textit{Is the
    universe open or closed? The density of matter in the universe} Cambridge
    Lecture Notes in Physics, vol. 7. (Cambridge University Press,
    Cambridge 1997).


\bibitem{Einstein:1937} A.~Einstein, and N.~Rosen,
\newblock {Journal of Franklin Institute} {\bf 223}, 43--54 (1937).

\bibitem{Kennefick05} D.~Kennefick,
\newblock {Physics Today} {\bf 58}, 43 (2005).

\bibitem{Bondi:1957dt} H.~Bondi,
\newblock {Nature} {\bf 179}, 1072--1073 (1957).

\bibitem{Pirani:1956wr} F.~A.~E.~Pirani,
\newblock {Phys.\,Rev.} {\bf 105}, 1089--1099 (1957).

\bibitem{Bondi:1959} H.~Bondi, F.~A.~E.~Pirani and
    I.~Robinson, {Proc.\,Roy.\,Soc.\,Lond.} {\bf A251}, 519 (1959).

\bibitem{isaacson68a} R.~A.~Isaacson,
\newblock {Phys. Rev.} {\bf 166}, 1263 (1968).

\bibitem{Isaacson68b} R.~A.~Isaacson,
\newblock {Phys. Rev.} {\bf 166}, 1272 (1968).

\bibitem{Schwarzschild16} K.~Schwarzschild,
\newblock {Sitzber. Deut. Akad. Wiss. Berlin, Kl.
Math-Phys. Tech.} 189 (1916).

\bibitem{Kerr63} R.~P.~Kerr,
\newblock {\em Phys. Rev. Lett.}, {\bf 11}, 237 (1963).

\bibitem{Oppenheimer39a} J.~R.~Oppenheimer, and G.~M.~Volkoff,
         Phys. Rev. {\bf 55}, 374 (1939).

\bibitem{Oppenheimer39b} J.~R.~Oppenheimer and H.~Snyder,
\newblock {Phys. Rev.} {\bf 56}, 455 (1939).

\bibitem{May66} M.~M.~May and R.~H.~White,
\newblock {Phys. Rev.} {\bf 141}, 1232 (1966).

\bibitem{Kennefick07} D.~Kennefick,
\newblock {\em Traveling at the Speed of Thought: Einstein and the Quest for
  Gravitational Waves}, (Princeton University Press, Princeton 2007).

\bibitem{Weber57} J.~Weber and J.~A.~Wheeler,
\newblock {Rev. Mod. Phys.} {\bf 29}, 509--515 (1957).

\bibitem{Weber67} J.~Weber,
\newblock {Phys. Rev. Lett.} {\bf 18}, 498 (1967).

\bibitem{Weber69} J.~Weber,
\newblock {Phys. Rev. Lett.} {\bf 12}, 1320--1324 (1969).

\bibitem{saulson05} P.~Saulson, In: {\em Receiving
    Gravitational Waves}, (World Scientific, Singapore 2005),
    Chap.~9.

\bibitem{Aguiar11} O.~D.~Aguiar,
\newblock {Res. Astron. Astrophys.} {\bf 11}, 1--42 (2011).

\bibitem{Levine72} J.~Levine and R.~Stebbins,
\newblock {Phys. Rev. D} {\bf 6}, 1465--1468 (1972).

\bibitem{Giganti77} J.~J.~Giganti, J.~V.~Larson, J.~P.~Richard,
    R.~L.~Tobias, and J.~Weber,
\newblock {\em Lunar Surface Gravimeter Experiment Final Report} (1977).
\newblock See
  http://www.lpi.usra.edu/lunar/missions/apollo/apollo\_17/experiments/lsg/.

\bibitem{Armstrong06} J.~W.~Armstrong,
\newblock {Living Rev. Relativity} {\bf 9}, 1 (2006).

\bibitem{saulsonbook} P.~Saulson,
\newblock {Fundamentals of Interferometric Gravitational Wave
Detectors}, (World Scientific, Singapore (1994).

\bibitem{Braginsky77} V.~B.~Braginsky, Yu.~I.~Vorontsov,
    and F.~Ya.~Khalili,
\newblock {Sov. Phys. JETP} {\bf 46}, 705 (1977).

\bibitem{Caves81} C.~Caves,
\newblock {Phys. Rev. D.} {\bf 23}, 1693--1708 (1981).

\bibitem{Schnabel11} R.~Schnabel et~al.
\newblock {Nature Physics Letter} {\bf 7}, 962--965 (2011).

\bibitem{Aasi13b} J.~Aasi et~al,
\newblock {Nature Photonics Letter} {\bf 7}, 613--619 (2013).

\bibitem{Smarr79} L.~L.~Smarr,
\newblock {Gauge Conditions, Radiation Formulae and the Two Black Hole
  Collision}, (Cambridge University Press, Cambridge 1979),
\newblock Page  245.

\bibitem{Hulse75} R.~A.~Hulse and J.~H.~Taylor,
\newblock {Astrophys. J. Lett.}, {\bf 195}, L51 (1975).

\bibitem{Abadie:2010cfa} J.~Abadie et~al,
\newblock {Class. Quant. Grav.} {\bf 27}, 173001 (2010).

\bibitem{Hawking71} S.~W.~Hawking,
\newblock {Phys. Rev. Lett.} {\bf 26}, 1344--1346 (1971).

\bibitem{Pretorius2005a} F.~Pretorius,
\newblock {Phys.\,Rev.\,Lett.} {\bf 95}, 121101 (2005).

\bibitem{Baker2006a} J.~G.~Baker, J.~Centrella, D.~Choi,
    M.~Koppitz, and J.~van Meter,
\newblock {Phys.\,Rev.\,Lett.} {\bf 96}, 111102 (2006).

\bibitem{Campanelli2006a} M.~Campanelli, C.~Lousto,
    P.~Marronetti and Y.~Zlochower,
\newblock {Phys.\,Rev.\,Lett.} {\bf 96}, 111101 (2006).

\bibitem{Allen99} B.~Allen et~al,
\newblock {Phys. Rev. Lett.} {\bf 83}, 1498 (1999).

\bibitem{Abadie12} J.~Abadie et~al,
\newblock {Sensitivity Achieved by the {LIGO} and {V}irgo Gravitational
  Wave Detectors during {LIGO}'s Sixth and {V}irgo's Second and Third Science
  Runs}, \newblock arXiv:1203.2674 [gr-qc] (2012).

\bibitem{Astone10} P.~Astone et~al,
\newblock {Phys. Rev. D} {\bf 82}, 022003 (2010).

\bibitem{Abbott07} B.~Abbott, et~al,
\newblock {Phys. Rev. D} {\bf 76}, 022001 (2007).

\bibitem{Hobbs10} G.~Hobbs et~al,
\newblock {Class. Quantum Grav.} {\bf 27}, 084013 (2010).

\bibitem{BICEP14} P.~A.~R.~Ade et~al,
\newblock {Phys. Rev. Lett.} {\bf 112}, 241101 (2014).

\bibitem{Sathya09} B.~S.~{Sathyaprakash} and {B.~F.~Schutz},
\newblock {Living Rev. Relativity} {\bf 12}, 2 (2009).

\bibitem{Will14} C.~M.~Will,
\newblock {Living Rev. Relativity} {\bf 17}, 4 (2014).

\bibitem{Punturo10} M.~Punturo et~al,
\newblock {Class. Quantum Grav.} {\bf 27}, 194002 (2010).


\bibitem{CB52}  Y. Choquet--Bruhat, Acta Math. \textbf{88} 141-225
    (1952).

\bibitem{L44}  A. Lichnerowicz, J. Math. Pures Appl. \textbf{23}
    37-63 (1944).

\bibitem{Y72} J. York, J. Math. Phys. \textbf{13} 125-130 (1972).

\bibitem{HNT} M. Holst, G. Nagy and G. Tsogtgerel,  Comm. Math.
    Phys. \textbf{288} 549-613 (2009).

\bibitem{Max11} D. Maxwell, Comm. Math. Phys. \textbf{302} 697-736
    (2011).

\bibitem{Y99} J. York, Phys. Rev. Lett. \textbf{82} 1350-1353
    (1999).

\bibitem{M14a} D. Maxwell, unpub. gr-qc 1402.5585.

\bibitem{M14b} D. Maxwell, unpub. gr-qc 1407.1467.

\bibitem{IMP02} J. Isenberg, R. Mazzeo and D. Pollack, Comm. Math.
    Phys. \textbf{231}  529-568 (2002).

\bibitem{CIP} P. Chrusciel, J. Isenberg and D. Pollack, Comm. Math.
    Phys. \textbf{257}  29-42 (2005).

\bibitem{C00} J. Corvino, Comm. Math. Phys. \textbf{214} 137-189
    (2000).

\bibitem{CS06} J. Corvino and R. Schoen, J. Diff. Geom. \textbf{73}
    185-217 (2006).

\bibitem{CS14} A. Carlotto and R. Schoen, unpub. math 1407.4766.

\bibitem{CCI11} P. Chrusciel, J. Corvino and J. Isenberg, Comm.
    Math.Phys. \textbf{304} 637-647 (2011).

\bibitem{B93} R. Bartnik, J. Diff. Geom. \textbf{37} 31-71, (1993).

\bibitem{BI06} R. Bartnik and J. Isenberg  Class. Qtm. Grav.
    \textbf{23} 2559-2569 (2006).

\bibitem{ADM} R. Arnowitt, S. Deser and C. Misner, Phys. Rev.
    \textbf{122} 997-1006 (1961).

\bibitem{SY79} R. Schoen and S.--T. Yau, Comm. Math. Phys.
    \textbf{65} 45-76 (1979).

\bibitem{W81} E. Witten, Comm. Math. Phys. \textbf{80} 381-402
    (1981)

\bibitem{HI01} G. Huisken and T. Ilmanen, J. Diff. Geom. \textbf{59}
    353-437 (2001).

\bibitem{B01} H. Bray, J. Diff. Geom. \textbf{59} 177-267 (2001).

\bibitem{EHLS} M. Eichmair, L.--H. Huang, D. Lee and R. Schoen,
    unpub. math 1110.2087.

\bibitem{Ger} R. Geroch, J. Math. Phys. \textbf{9} 450-465 (1968).

\bibitem{CBG} Y. Choquet--Bruhat and R. Geroch, Comm. Math. Phys.
    \textbf{14} 329-335 (1968).

\bibitem{Sb} J. Sbierski, unpub. gr-qc 1309.7591.

\bibitem{Mon81} V. Moncrief, Phys. Rev. D \textbf{23} 312-315
    (1981).

\bibitem{Ring09} H. Ringstr\"om,  Ann. Math.\textbf{170} 1181-1240
    (2009).

\bibitem{CK94} D. Christodoulou and S. Klainerman, \textit{The
    Global Nonlinear Stability of Minkowski Space}, Princeton U.
    Press (1994).

\bibitem{Chr09} D. Christodoulou, \textit{The Formation of Black
    Holes in General Relativity}, EMS (2009).

\bibitem{F86} H. Friedrich,  J. Geom. Phys. \textbf{3} 101-117
    (1986).

\bibitem{AM04} L. Andersson and V. Moncrief, in \textit{Cauchy
    Problem in General Relativity}, ed. P. Chrusciel and H.
    Friedrich, Birkahauser (2004).

\bibitem{Ring08} H. Ringstr\"om, Invent. Math. \textbf{173} 123-208
    (2008).

\bibitem{Ring09a} H. Ringstr\"om,  Comm. Math. Phys. \textbf{290}
    155-218 (2009).

\bibitem{Sv} C. Svedberg, Ann H. Poinc. \textbf{12} 849--917 (2011).

\bibitem{LI13} X. Luo and J. Isenberg, Ann. Phys. \textbf{334}
    420-454 (2013).

\bibitem{KRS} S. Klainerman, I. Rodnianski and J. Szeftel, unpub.
    gr-qc 1204.1772.

\bibitem{EM82} D. Eardley and V. Moncrief, Comm. Math. Phys.
    \textbf{83}, 171-212 (1982).

\bibitem{Pr05} F. Pretorius, Phys. Rev. Lett. \textbf{95} 121101
    (2005).

\bibitem{HL64} S. Hahn and R. Lindquist, Ann. Phys. \textbf{29}
    304-331 (1964).

\bibitem{Ca06} M. Campanelli et. al., Phys. Rev. Lett. \textbf{96}
    111101 (2006).

\bibitem{Ce06} J. Baker et. al., Phys. Rev. Lett. \textbf{96} 111102
    (2006).

\bibitem{Ch94} M. Choptuik, Phys. Rev. Lett. \textbf{70} 9-12
    (1994).

\bibitem{BM93} B. Berger and V. Moncrief, Phys. Rev. D \textbf{48},
    4676-4687 (1993).

\bibitem{BKL} V. Belinskii, E. Lifshitz, and I.  Khalatnikov,  Zh.
    Eksp. Teor. Fiz. \textbf{62}, 1606-1613 (1972).

\bibitem{Ring09b} H. Ringstr\"om,  Ann. Math. \textbf{170}
    1181-1240 (2009).

\bibitem{Biz14} P. Bizon, Gen. Rel. Grav. \textbf{46} 1724 (2014).

\bibitem{LP10} L. Lehner and F. Pretorius, Phys. Rev. Lett.
    \textbf{105} 101102 (2010).







\bibitem{hayward}S.~A.~Hayward, \texttt{gr-qc/0008071}.

\bibitem{aaetal} A.~Ashtekar et al, Phys. Rev. Lett.
    \textbf{85}, 3564-3567 (2000).

\bibitem{akrev} A.~Ashtekar and B.~Krishnan, Living Rev. Relativity
    \textbf{7} 10 (2004).

\bibitem{gjrev} E.~Gourgoulhon, J.~L.~Jaramillo, Phys. Rept.
    \textbf{423} 159-294 (2006).

\bibitem{ac} A.~Ashtekar and M.~Campiglia, Phys. Rev. D\textbf{88},
    064045 (2013).

\bibitem{ro} R.~Owen, Phys. Rev. D{\bf 80}, 084012 (2009).

\bibitem{bch} J.~M.~Bardeen, B.~Carter and S.~W.~Hawking, Commun.
    Math. Phys. \textbf{31}, 161-170 (1973).

\bibitem{leshouches}\textit{Black Holes, Les Houches 1972}, C.
    DeWitt and B. DeWitt eds (Gordon and Breach, New York, 1973).

\bibitem{apr} A.~Ashtekar, F.~Pretorius and F.~M. Ramazanoglu, Phys.
    Rev. Lett. \textbf{106} 161303 (2011); Phys. rev. D\textbf{83},
    044040 (2011).

\bibitem{core} J.~M.~M Senovilla, In {\it Relativity and
    Gravitation: 100 Years after Einstein in Prague}, J. Bicak and
    T.~Ledvinka, eds (Springer, Berlin, 2014).

\bibitem{ag} A.~Ashtekar and G.~J.~Galloway, Adv. Theor. Math.
   Phys. {\bf 9} 1-30 (2005).

\bibitem{gr3}  \textit{Proceedings on theory of gravitation},
    I.~Infeld ed (Guthier Villars, Paris and PWN Editions Scientific
    de Pologne, Warszawa (1964)).




\bibitem{ae}A.~Einstein, \textit{Meaning of Relativity}
    (Princeton University Press, Princeton, 1945).

\bibitem{rosenfeld} L.~Rosenfeld, Ann. Physik \textbf{5}, 113
    (1930); Z. Physik \textbf{65} 587 (1930).

\bibitem{bron1}M.~P.~Bronstein, Phys. Zeitschr. der Sowjetunion
    \textbf{9}, 140-157 (1936);\\
     S. Deser and A. Starobinski, Gen. Rel. Grav. \textbf{44} 263-265 (2012).

\bibitem{bron2} M.~P.~Bronstein, Zh.ETF (JETP) \textbf{6} 195
    (1936);\\
    D.~J.~Cirlio-Lombardo, arXiv:1405.2334.

\bibitem{rpf}R.~P.~Feynman, Acta Physica Polonica, \textbf{24},
    697-722 (1964).

\bibitem{bsd}B.~S.~DeWitt, in \textit{Magic without Magic: John
    Archibald Wheeler} edited by J.~R. Klauder (W. H. Freeman,
    San Fransico 1972).

\bibitem{thv} G. 't~Hooft and M.~J.~G. Veltman, Annales Poincar\'e
    Phys. Theor. \textbf{A20} 69 (1974).

\bibitem{gs}M.~H. Goroff and A.~Sagnotti, Nucl. Phys.
    \textbf{B266} 709 (1986).

\bibitem{asymfree} K.~S.~Stelle, Phys. Rev. D\textbf{16} 953
    (1977);\\
    T.~Tomboulis, Phys. Lett. B\textbf{97} 77 (1980).

\bibitem{sugra1} S.~Deser, J.~Kay and K.~S.~Stelle, Phys. Rev. Lett.
\textbf{38}, 527 (1977); \\
M.~T.~Grisaru, P. Van Nieuwenhuizen and J.~A.~M.~Vermaseren, Phys.
Rev. Lett. \textbf{37}, 1662 (1976).

\bibitem{howe-stelle} P.~S. Howe and K.~S.~Stelle, Int. J. Mod.
    Phys. A\textbf{4}, 1871 (1989).

\bibitem{parker} L.~Parker, Ph. D. Dissertation (1967).

\bibitem{adm} R.~Arnowitt, S.~Deser and C.~W.~Misner, 
    in \textit{Gravitation: An introduction to current research}
    ed L.~Witten (John Wiley, New York, 1962).

\bibitem{komar} A.~Komar,
    in \textit{Relativity} Carmeli M, Fickler S. I. and
    Witten L (eds) (Plenum, New York, 1080).

\bibitem{pbak} P.~G.~Bergmann and A.~Komar, 
    \textit{General Relativity and
    Gravitation vol 1, On Hundred Years after the Birth of Albert
    Einstein}, ed A.~Held  (Plenum, New York, 1980).

\bibitem{agrev} A.~Ashtekar and R.~Geroch, {Quantum theory of
    gravitation}, \textit{Rep. Prog. Phys.} \textbf{37} 1211-1256
    (1974).

\bibitem{kk1} K.~Kucha$\check{\rm r}$, 
    in  \textit{Quantum Gravity 2, A Second Oxford
    Symposium} ed C.~J.~Isham, R.~Penrose and D.~W.~Sciama (Clarendon
    Press, Oxford, 1975).

\bibitem{jw1} J.~A.~Wheeler,{\it Geometrodynamics}, (Academic
    Press, New York, 1964).

\bibitem{jw2} J.~A.~Wheeler, 
     in \textit{Relativity, Groupos and Topology} eds DeWitt
    C M and DeWitt B S (Gordon and Breach, New York, 1963).

\bibitem{weinberg} S.~Weinberg \emph{Gravitation and
    Cosmology} (John Wiley, New York, 1972).

\bibitem{casimir} L.~H.~Ford and C.~H.~Wu, AIP Conf. Proc.
    \textbf{977} 145-159 (2008).

\bibitem{timemachines} A.~Everett and T.~Roman, \emph{ Time Travel
    and Warp Drives: A Scientific Guide to Shortcuts through Time
    and Space} (U of Chicago, Chicago, 2012).

\bibitem{effective} C.~P.~Burgess, Living Rev. Relativity \textbf{7}
    5 (2004).

\bibitem{rothstein} R.~A.~Porto and I.~Z.~Rothstein, Phys. Rev.
    D\textbf{78} 044013 (2008), D\textbf{81} 029902 (2010).

\bibitem{asymquan}A. ~Ashtekar, Phys. Rev. Lett. \textbf{46}, 573
    (1981); J. Math. Phys. \textbf{22} 2885 (1987);
\textit{Asymptotic Quantization} (Bibliopolis, Naples, 1987); \\
A.~Strominger, {arXiv:1312.2229}\\
T.~He, V.~Lysov, P.~Mitra and A.~Strominger, {arXiv:1401.7026}.

\bibitem{scattering} L.~Mason and D.~Skinner, {arXiv:0808.3907};\\
 F.~Cachazo and Y.~Geyer, arXiv:1206.6511;\\
F.~Cachazo, S.~He, and Y.~E.~Yuan., {arXiv:1307.2199};\\
N.~Arkani-Hamed and J.~Trnka, {arXiv:1312.2007}.

\bibitem{rpwr}R.~Penrose and W.~Rindler, \textit{Spinors and
    Space-Time: Volume 2, Spinor and Twistor Methods in Space-Time
    Geometry} (Cambridge University Press, Cambridge 1988).

\bibitem{hamber}H.~W.~Hamber, \textit{Quantum gravitation: The
    Feynman path integral approach} (Springer, Berlin (2009)).

\bibitem{horava} {P.~Ho\v{r}ava,
Phys. Rev. D\textbf{79} 084008 (2009).}

\bibitem{causets} F.~Dowker, Gen. Rel. Grav. \textbf{45}, 1651-1667
    (2013).

\bibitem{vassiliev} M.~Vassiliev, \textit{Introduction to Higher Spin
    Gauge Theory}, Kramer's course, (University of Utretcht, Utretcht
    (2014)).

\bibitem{wolf}\textit{Some Strangeness in
    Proportion}, H.~Wolf, ed. (Addison Wesley, Reading, 1980).

\bibitem{hi} \textit{General Relativity, an Einstein Centennial
    survey} S.~W.~Hawking and W.~Israel, eds. (Cambridge University
    Press, Cambridge, 1979).

\bibitem{swh1} S.~W.~Hawking, \textit{Is the End In Sight for
    Theoretical Physics?: An Inaugural Address} (Cambridge University
    Press, Cambridge, 1980).

\bibitem{carlip}S.~Carlip, {arXiv:1207.4503}.

\bibitem{nicolai}H.~Nicolai, Physics \textbf{2}, 70 (2009); Personal
    communication (2014).

\bibitem{gross-periwal}D.~J.~Gross and V.~Periwal, Phys. Rev. Lett.
    \textbf{60} 1517 (1988).

\end{thebibliography}

\end{document}